\documentclass[10pt,journal,compsoc]{IEEEtran}

\ifCLASSOPTIONcompsoc
\usepackage[nocompress]{cite}
\usepackage[noend]{algorithmic}
\usepackage{caption}
\usepackage{graphicx}
\usepackage{booktabs}
\usepackage{amsmath}
\usepackage{amsthm}
\usepackage{multirow}
\usepackage{amsmath}
\usepackage[ruled,vlined,linesnumbered]{algorithm2e}
\usepackage{diagbox}
\usepackage{color}
\usepackage{amstext}
\usepackage{booktabs}
\usepackage{ragged2e}
\usepackage{amssymb}
\usepackage{multicol,algorithmic}
\usepackage{listings}
\usepackage{multirow}
\usepackage{float}

\else
\usepackage{cite}
\fi

\ifCLASSINFOpdf

\else

\fi

\begin{document}

\title{A proof of contribution in blockchain using game theoritical deep learning model}
%
\author{Jin Wang,~\IEEEmembership{IEEE Senior Member}
        \IEEEcompsocitemizethanks{\IEEEcompsocthanksitem Jin Wang is with the College of Internet of Things Engineering, Wuxi University, Wuxi 214105, China. (sophie\_icon@sina.com)}%
        \thanks{Manuscript received \today; revised ~~~~~~~~~..}%
}

\markboth{Game-Theoretic Deep Learning Models}%
{Shell \MakeLowercase{\textit{et al.}}: Bare Demo of IEEEtran.cls for Computer Society Journals}

\IEEEtitleabstractindextext{%
\justifying\let\raggedright\justifying
\begin{abstract}
Building elastic and scalable edge resources is an inevitable prerequisite for providing platform-based smart city services. Smart city services are delivered through edge computing to provide low-latency applications.
However, edge computing has always faced the challenge of limited resources.
A single edge device cannot undertake the various intelligent computations in a smart city, and the large-scale deployment of edge devices from different service providers to build an edge resource platform has become a necessity.
Selecting computing power from different service providers is a game-theoretic problem.
To incentivize service providers to actively contribute their valuable resources and provide low-latency collaborative computing power, we introduce a game-theoretic deep learning model to reach a consensus among service providers on task scheduling and resource provisioning.
Traditional centralized resource management approaches are inefficient and lack credibility, while the introduction of blockchain technology can enable decentralized resource trading and scheduling.
We propose a contribution-based proof mechanism to provide the low-latency service of edge computing.
The deep learning model consists of dual encoders and a single decoder, where the GNN (Graph Neural Network) encoder processes structured decision action data, and the RNN (Recurrent Neural Network) encoder handles time-series task scheduling data.
Extensive experiments have demonstrated that our model reduces latency by $5.84\%$ compared to the state-of-the-art.
\end{abstract}
}

\maketitle

\IEEEdisplaynontitleabstractindextext

\IEEEpeerreviewmaketitle

\IEEEraisesectionheading{\section{Introduction}\label{sec:introduction}}
\IEEEPARstart {E}{dge} computing involves a large number of distributed resources, including computation, storage, and bandwidth, which require efficient coordination and management of these dispersed resources.
The traditional centralized resource management approach is inefficient and lacks credibility, while the introduction of blockchain technology can realize decentralized resource trading and scheduling, significantly improving the transparency, efficiency, and resource utilization of transactions.
The common consensus mechanisms include Proof-of-Work (PoW) and Proof-of-Stake (PoS), but they are relatively one-sided and do not sufficiently require participants to contribute to the community.
Additionally, in edge computing, there is a lack of horizontal collaboration among geographically distributed edge nodes, which limits the coverage of service areas.

\begin{figure}
\centering
\includegraphics[scale=0.41]{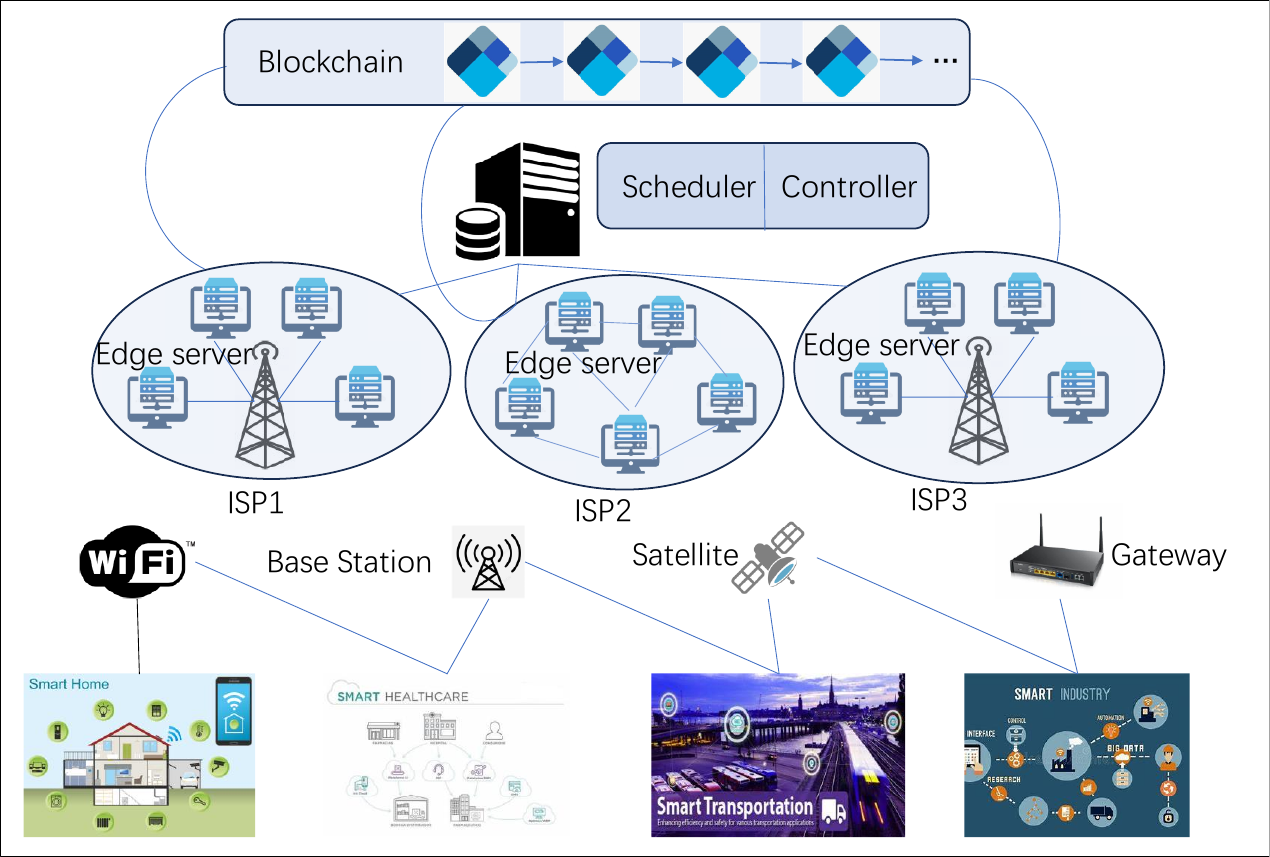}
\caption{Blockchain-based collaborative edge network}
\label{figure00}
\end{figure}
In the reference \cite{f4}, a blockchain-based collaborative edge intelligence (BCEI) approach can be adopted in a collaborative edge network.
As shown in Figure \ref{figure00}, geographically distributed edge devices form a peer-to-peer network and maintain a permissioned blockchain, with the results written immutably to the blockchain to ensure reliability.
Through this edge-to-edge collaboration mode, data and computational resources are shared to execute computationally intensive tasks.
The edge nodes belong to different stakeholders.
The process of reaching transaction consensus on the edge resource trading blockchain is the selection of which device vendors will serve as resource providers.
Game theory can generate consensus knowledge, and nodes must choose strategies to reach consensus and decide on blockchain state updates.

We propose a PoC (proof-of-contribution) consensus mechanism, which selects the resource inputs participating in the computation and competition to maximize the contribution to latency reduction.
PoC determines the influence of participants in the network based on their computing capabilities and resource levels, which has a positive impact on the performance efficiency of the edge network.
The PoC consensus mechanism has the characteristics of low energy consumption and high transaction throughput in terms of scalability, and the transaction time and efficiency are not negatively impacted as the number of participants increases.
The blockchain consensus mechanism is a dynamic game process in which participants reach consensus through certain rules and incentive mechanisms.

On heterogeneous platforms for edge resources, the collaboration between different device vendors is also a game.
The goal of this game system is not merely to bring the maximum return to the participants, but to bring the greatest benefit to the entire edge computing community.
There are deep connections between game theory and deep learning.
On the one hand, deep learning can be represented as a game process due to its dynamic interactive environment.
For example, most of the classification problems that commonly employ deep learning approaches can be seen as a Stackelberg game \cite{f1}.
On the other hand, game theory problems can be solved using deep learning, because deep learning networks have powerful representation and learning capabilities.
By learning from historical data and environmental feedback, deep learning networks automatically adjust their network weights and parameters to obtain the optimal decisions.

The simple algorithms of consensus mechanisms may not be able to capture the complex data distributions, patterns, and player behaviors in a game, so the generated data may not fully reflect the diversity and variability of the real world.
Compared to traditional operational research methods, deep learning models can better handle large-scale, high-dimensional data and have stronger generalization capabilities.
Deep learning models can adapt to different game scenarios and learn more complex strategies and behavioral patterns by learning from a large amount of training data.
Meta-strategies can guide the game players' choices of actions in different situations, balancing exploration and exploitation, and how to respond to changes in opponent strategies.
Therefore, we propose to use GA (genetic algorithm) to generate meta-strategies and output them as training data, and then have the deep learning model learn the strategies and output actions by learning from the meta strategy training data.

Although multi-agent deep reinforcement learning can be used to solve the problems of collaboration and competition in multi-intelligent agent competition, its training complexity is high and it generally suffers from convergence difficulties.
Therefore, we use a dual encoder-decoder model, where each game participant learns based on the meta strategy of the jointly optimal strategy (for all edge nodes), so that the service latency performance of the resource alliance provided to the outside is optimized.
This is a multi-party cooperative game, where only one person can obtain each job, and multiple players cooperate in the game process to achieve the common goal.
Players with lower resource utilization are easier to obtain tasks at the beginning, but when that player has obtained more tasks, the wait time in the resource queue will become longer, at which point the tasks will start to shift to other players.

In summary, the main contributions of this paper are as follows:
\begin{itemize}
\item[\textbullet]We propose a PoC consensus mechanism, which selects the resource input and task resource providers participating in the computation and competition, in order to maximize the contribution to the community.
    This contribution takes into account the minimization of the average task latency.
    All community resource transactions are recorded on the blockchain.
    From the perspectives of efficiency and fairness, the qualification for on-chain transactions is granted based on the contribution of the game participants to the entire edge computing community.
\item[\textbullet]
    We adopt a dual encoder (GNN encoder+RNN encoder) and a single decoder structure, where the two encoders are fused to obtain a comprehensive hidden representation, extracting the integrated features between task information and time series information.
    The decoder is used to convert the learned hidden representation into the actual resource allocation plan or game decision.
    This model can simultaneously consider global and local features, thereby better modeling the strategy generation process in the game.
    Forming and optimizing the time series of task scheduling, the deep learning model often has good diversity and robustness by relying on or leveraging the meta-strategy, a higher-level guidance mechanism, and can enhance the model's generalization ability in complex environments.
\item[\textbullet]The diverse game behavior simulated by deep learning networks means that even if some nodes fail or behave maliciously, the entire system can still function normally.
    The diversity of game behavior aligns with the core idea of Byzantine fault tolerance, allowing the blockchain system to maintain secure and stable operation.
    Each participant is unable to fully understand the internal state of the other participants, which reduces the likelihood of the system being disrupted.
\item[\textbullet]Since edge nodes may belong to different stakeholders, centralized management can lead to issues of trustworthiness and privacy.
    Therefore, we adopt a knowledge distillation approach to deploy the PoC consensus mechanism, where the meta strategy model and resource allocation teacher model are maintained by the resource management system/center, while the resource allocation student model learns on the distributed edge nodes.
    The student models achieve on-chain PoC transactions of resources through peer-to-peer communication and decentralized consensus among themselves.
\end{itemize}

The remaining sections of this paper are organized as follows: Section 2 reviews related work in blockchain-based edge computing and game.
Section 3 describes the problem formalization and modeling.
Section 4 presents our proposed method, providing detailed constructions of the game theoritical deep learning model.
Section 5 describes the experimental setup and evaluation results.
Finally, in Section 6, we summarize the contributions of this paper and discuss future research directions in the field.
\section{Related works}\
The current mainstream blockchain consensus mechanisms include: PoW (Proof of Work), PoS (Proof of Stake), DPoS (Delegated Proof of Stake), and PBFT (Practical Byzantine Fault Tolerance) \cite{f7}.
In recent years, several other consensus mechanisms have emerged, such as PoB (Proof of Bandwidth)\cite{f8}, and PoA (Proof of Authority)\cite{f9}.
PoW requires a large amount of computational resources, while PoS is susceptible to manipulation by stakeholders.
The current consensus mechanisms do not contribute enough to the broader blockchain community.
The reference \cite{f10} views the consensus mechanism of the Bitcoin network as a typical Byzantine Generals' Problem, involving the game-theoretic behavior of the participating nodes.

It is widely recognized that many dynamically interactive environments can be represented as games.
Game theory can be used to analyze the strategies of the consensus nodes as well as the interactions among them\cite{f11}.
The core of deep learning is referred to as a game problem, where individual components are each pursuing their own optimal solutions while also being influenced by the other components.
The reference \cite{f6} discusses the interplay between deep learning and game theory.
It models basic deep learning tasks as strategic games.

Reinforcement learning is the combination of game theory and deep learning.
Multiple research papers have explored using deep reinforcement learning to implement game-playing, including the famous AlphaGo system, as well as examples like \cite{f12}, \cite{f13}, \cite{f14}.
However, when reinforcement learning learns in complex distributed environments, it requires a large amount of computational resources\cite{f15}.
While in multi-agent reinforcement learning, due to the complex interactions between agents, the convergence is poor, which affects the effectiveness of decentralized learning\cite{f16}.

Most of the classification problems which popularly employ a deep learning approach can be seen as a Stackelberg game \cite{f6}.
In classification problems, the model can be viewed as the Stackelberg leader, while the data is the follower.
In the resource allocation model we propose, the participants are on an equal footing, without any obvious leaders and followers.
The proposed model is more akin to a cooperative game framework, where the participants reach a win-win resource allocation scheme through coordination and gaming.

Game theory often assumes the rationality, complete information, and fixed strategy sets of the participants\cite{f2}.
However, in the real world, the decision-making of participants may be influenced by emotions, limited information, dynamically changing strategy choices, and other factors.
Therefore, the learning process simulated by deep model simulations of the real game process requires the diversity of game actions.
There is no fixed "state-of-the-art" solution or accuracy.
The diversity of game behavior is also an important consideration factor in the design of Byzantine fault-tolerant systems\cite{f7}.
The reason is that the diversity of the participants' actions is beneficial to avoid a single stakeholder controlling the entire system, and the diversity of consensus behavior will increase the cost for attackers.

Deep learning models often have thousands of layers in depth or hundreds of millions of parameters, requiring training on powerful GPU clusters \cite{f32}.
In edge computing networks, edge devices have very limited resources, so knowledge distillation is introduced to compress and accelerate the models.
Another reason for deploying knowledge distillation is the lack of credibility and potential privacy leakage issues in the centralized deployment of blockchain consensus mechanism models \cite{f4}.
Due to the complexity of multi-agent environments and the need for agents to balance exploration and exploitation \cite{f33}, knowledge distillation on reinforcement learning models like MADDPG (Multi-Agent Deep Deterministic Policy Gradient) is much more challenging than on regular deep models.
Initially proposed in \cite{f34}, the core idea of knowledge distillation is to learn a small "student" model from a large "teacher" model.
The teacher model can arbitrate the decision discrepancies of the student model when necessary, and the student models can communicate with each other to eventually reach a decentralized consensus.
The iterative optimization and strategy adjustment in knowledge distillation is similar to the best response process in game theory.
Therefore, this paper adapts game models to knowledge distillation systems, leveraging the powerful modeling capability and intelligent decision-making strategy of deep game models to better interact with the complex student models.

Autoencoder, as an unsupervised neural network model, has long been used to compress input data.
Recent research, such as \cite{f28} and \cite{f29}, has introduced additional Transformer encoders in the encoding stage of the encoder-decoder architecture, providing extra global modeling capabilities.
These models are often used for tasks like text translation, where multiple-head attention mechanisms are incorporated to enhance the model's ability to capture key information \cite{f30}.
Another extension of the autoencoder is the two-encoder-one-decoder architecture, which is commonly used for multi-source sequence-to-sequence learning tasks.
In literature \cite{f31}, the two encoders use BiLSTM (Bidirectional Long Short-Term Memory) and Transformer to encode the input text respectively, and the single decoder leverages attention mechanisms to fuse the outputs of the two encoders and generate the target language translation.
The proposed dual-encoder single-decoder model uses a GNN encoder to process structured decision action data, and an RNN encoder to handle time-series task scheduling data.

Existing research on deep learning meta-strategies includes:
1. Hyperparameter optimization, such as Random Search \cite{f17}, Bayesian Optimization \cite{f18}, and Hyperband \cite{f19}.
2. Transfer learning, such as Domain Adaptation\cite{f20}, Meta-Learning\cite{f21}.
3. Network architecture search, such as AutoML (Automated Machine Learning) \cite{f22}, NAS (Neural Architecture Search) \cite{f23}, and DARTS (Differentiable Architecture Search) \cite{f24}.
4. Model ensembling: such as Boosting\cite{f25}, Bagging\cite{f26}, and Stacking\cite{f27} ensemble learning methods.
The proposed meta-strategy is a high-level decision process, which can obtain the optimal training data and improve the overall performance of the model.
Meta strategy often has good diversity and robustness, which can enhance the model's generalization ability in complex environments.
\section{Model}
We define four PyTorch model classes: GNNModel, RNNModel, ResourceAllocationModel, and PolicyModel.
The task characteristics representation captures the interactions and dependencies between multiple players, and can be represented as a graph structure.
The GNN encoder can capture the relationships between nodes in the graph.
GNN updates the nodes by aggregating the information of the task nodes, thereby integrating the global and local context.
This is particularly important for game tasks, as the interactions between players may involve both global strategy and local situation analysis.
The RNN encoder can handle and predict time series.

Here is a brief introduction to each model:
\begin{itemize}
\item[\textbullet]GNNModel:
This is a graph neural network-based model, containing a graph convolutional layer and a fully connected layer.
The input includes a graph structure g and node features x, and the output is node embeddings.
\item[\textbullet]RNNModel:
This is a recurrent neural network model, containing an RNN layer and a fully connected layer.
The input is sequence data x, and the output is the final hidden state.
\item[\textbullet]ResourceAllocationModel:
This is a composite model that combines the above two model structures.
The input includes graph structure gnn-input and sequence data rnn-input, and the output is a time series prediction.
Internally, the model uses GNNModel and RNNModel as encoders to encode the input into feature vectors.
Then it uses a linear layer and a Sigmoid activation function as the decoder to generate the final time series prediction.
This model aims to leverage the characteristics of graph structures and sequence data, and through the combination of graph neural networks and recurrent neural networks, to perform time series prediction for resource allocation. This hybrid model structure may be able to better capture the complex patterns and correlations in the input data.
\item[\textbullet]Policy-Model:
The policy model can act as a meta strategy to generate corresponding training datasets, for the ResourceAllocationModel to learn resource allocation strategies.
\end{itemize}

Objective function: Minimize latency
\begin{equation}
f_(x) = min\sum_{i=1}^n \max(0, C_i - a_i)
\end{equation}
\\Where $x = (C_1, C_2, ..., C_n)$ represents the completion times of the tasks.
\\$n$ represents the total number of tasks
\\$C_i$ represents the completion time of task $i$
\\$a_i$ represents the arrival time of task $i$
\\$\max(0, C_i - a_i)$ calculates the latency of task $i$, if the completion time is later than the arrival time, it is a positive value, otherwise it is 0.
\\$\sum_{i=1}^n \max(0, C_i - a_i)$ calculates the total latency of all tasks.
By optimizing the objective function by Genetic Algorithm (GA), we can generate a training sample set with the players' actions as labels, to provide the necessary samples for training the ResourceAllocationMod.
Through the data generation process, the policy model determines the overall resource allocation strategy, and then the various participants (followers) in the ResourceAllocationModel make optimal responses based on this.

For the consensus reached through game theory to obtain the right to transactions on the blockchain, the diversity of the moves chosen by the game participants can improve the diversity of the Byzantine fault-tolerant system.
The reason is that the diversity of the participants' actions is beneficial to avoid a single stakeholder controlling the entire system, and the diversity of consensus behavior will increase the cost for attackers.
We introduce the Euclidean distance diversity score. Assuming there are three game players, for each task the action set of each player is represented by a probability vector $P_i = (p_{i1}, p_{i2}, p_{i3})$, where $p_{i1}, p_{i2}, p_{i3}$ represent the probabilities of the three players being selected in action set $i$, respectively.
For each pair of action sets $P_i$ and $P_j$, we can calculate the distance between them as:
\begin{equation}
d(P_i, P_j) = sqrt[(p_{i1} - p_{j1})^2 + (p_{i2} - p_{j2})^2 + (p_{i3} - p_{j3})^2]
\end{equation}
To calculate the diversity score, we take the average of the Euclidean distances between all possible pairs of action sets (tasks). The formula is:
\begin{equation}
D = \frac 1 {(N*(N-1)/2)}\sum_{i=1}^N\sum_{j=i+1}^N d(P_i, P_j)
\end{equation}
Where N is the total number of action sets (tasks), and D is diversity score.
\section{Algorithm}
\subsection{Overview}
On the one hand, edge devices have limited resources and cannot deploy heavyweight deep learning models; on the other hand, the centralized deployment of blockchain consensus mechanism game models lacks credibility and is prone to privacy leaks.
In the knowledge distillation system, the teacher model enhances the decision-making capabilities of student models through knowledge transfer, and arbitrates the decision conflicts between students when necessary.
Student models can communicate with each other, and reach a decentralized consensus on the game actions that contribute the most to the blockchain based on the resource provision plans of the edge devices.

In terms of the specific models, the ResourceAllocationModel is a classification model with a fixed set of class labels, where each sample needs to be assigned to one of the classes.
However, this model cannot adequately represent the non-cooperative relationships between the participants, as well as their individual strategies and objective functions.
As a result, the model does not possess a Nash equilibrium solution.
In contrast, in cooperative game theory, the participants coordinate and negotiate with each other to collectively determine a strategy that maximizes the overall system's benefits.
While using deep learning models makes it difficult to simulate the negotiation process in gaming, it can still achieve the goal of benefit distribution in cooperative games.
Therefore, a more suitable approach would be to model the resource allocation problem as a cooperative game, where the participants work together to find an optimal solution that balances the two objectives of minimizing latency and balancing resource utilization.
This would allow the model to capture the strategic interactions and incentives of the various stakeholders, leading to a more realistic and effective resource allocation scheme.
\begin{figure}
\centering
\includegraphics[scale=0.66]{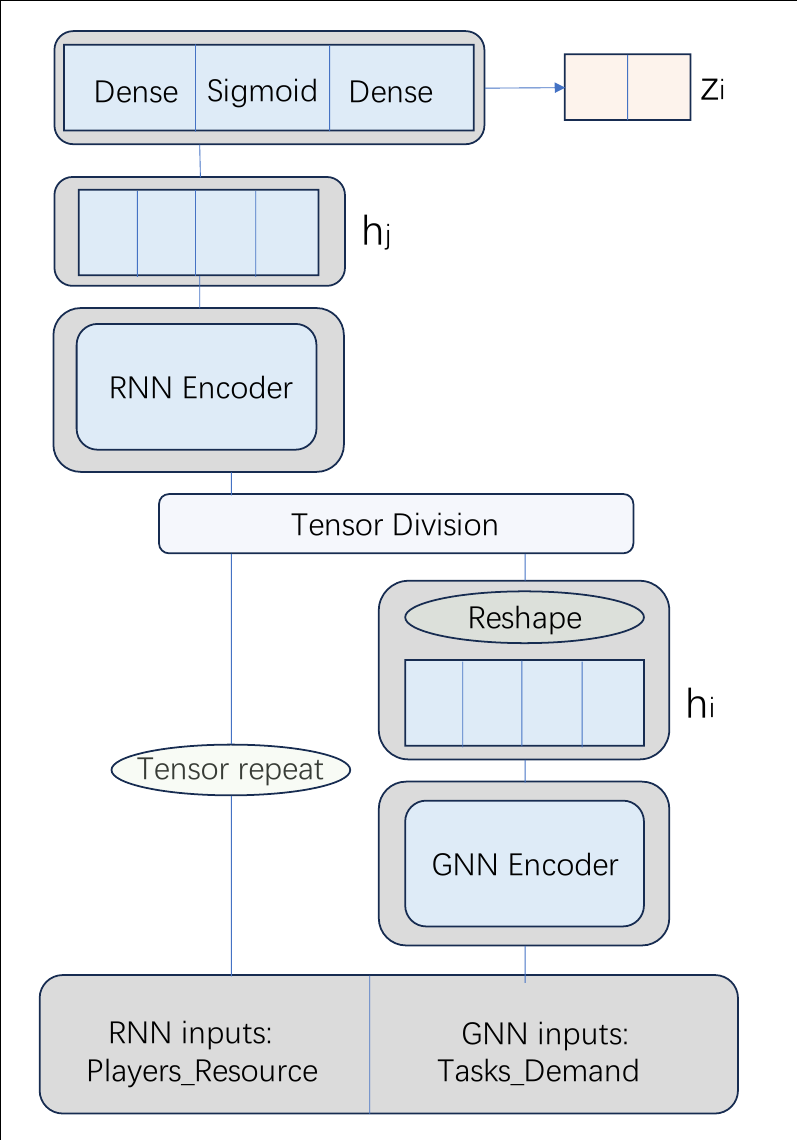}
\caption{Game theoritical deep learning model}
\label{figure01}
\end{figure}
\subsection{Dual-encoder and single decoder}
\begin{itemize}
\item[\textbullet]Using a combination of Graph Neural Networks (GNN) and Recurrent Neural Networks (RNN) to encode the input features can capture both the graph structure information and the time series information, which more comprehensively expresses the complexity of the resource allocation problem.
\item[\textbullet]By concatenating and processing the outputs of the GNN and RNN, the local graph information and the global time series information can be integrated to obtain a more balanced resource allocation scheme.
\item[\textbullet]Finally, using a linear layer and a Sigmoid function to generate the time series output, this output can be viewed as the resource allocation scheme among the participants, i.e., the benefit distribution in the cooperative game.
\item[\textbullet]During the training process, use the optimal solution generated by meta strategy that train the model by reducing latency as the target.
\end{itemize}
Integrating the points discussed above, the ResourceAllocationModel has to some extent achieved the goal of benefit distribution in cooperative game theory.
As shown in Figure \ref{figure01}, by establishing a joint model of GNN and RNN, and adopting an optimization strategy based on meta strategy, it has successfully transformed the resource allocation problem into a framework similar to cooperative game theory.
$h_i$, $h_j$ and $z_i$ in Figure \ref{figure01}, denote hidden features of two encoders, and the output.
The benefit of this approach is that it can not only obtain a fair resource allocation scheme that meets the needs of the various participants, but also consider the overall system performance metrics.
This allows the model to balance the individual interests of the participants as well as the collective optimization of the system as a whole.

The deep learning model adopts MSE Loss.
\begin{equation}
MSE Loss = \frac{1}{n} \sum_{i=1}^n(y_i-\hat{y}_i)^2
\end{equation}
$n$ is the number of samples, $y_i$ is the true value of the $i$-th sample, $\hat{y}_i$ is the predicted value of the $i$-th sample.
The MSE loss measures the average of the squared differences between the predicted values and the true values.
It is a commonly used loss function for regression tasks, which can effectively optimize the model to make the predicted values as close to the true values as possible.

\begin{figure*}
\centering
\includegraphics[scale=0.48]{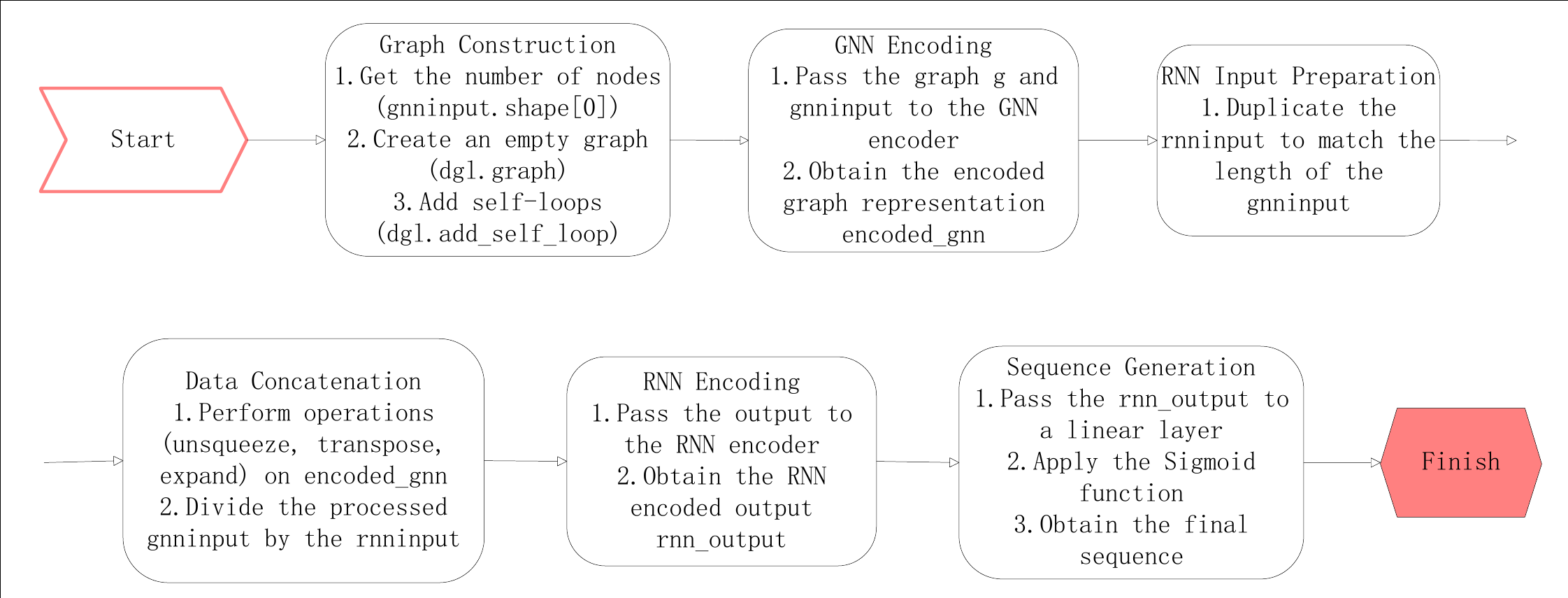}
\caption{Flowchat of Resource Allocation Model}
\label{figure02}
\end{figure*}
As shown in Figure \ref{figure02}, the flowchat of Resource Allocation model is divided into 6 parts: Graph Construction, GNN Encoding, RNN Input Preparation, Data Concatenation, RNN Encoding, and Sequence Generation.
This integrated approach of graph neural networks and recurrent neural networks can effectively capture the topological structure and temporal dynamics in resource allocation problems, thereby generating more optimized resource allocation schemes.
This is particularly important for resource supply game applications based on proof-of-contribution, as it can help achieve fair and reasonable resource allocation and consensus.
\subsection{Meta strategy}
The meta strategy of the deep learning model is to generate training data that reduces latency.
The problem is transformed into solving the single-objective optimization problem $\min f(x)$, where $f(x)$ represents the total task delay, and using the genetic algorithm to obtain the optimal solution $x^*$.
\\1. The objective function $f(x)$ is the total task delay, which can be calculated directly from the task allocation solution $x$.
This function does not have an analytical form of the gradient, so numerical optimization methods such as genetic algorithms are suitable for solving it.
\\2. The genetic algorithm is used to find the optimal task allocation solution $x^*$ that minimizes the total task delay $f(x)$.
A[Initialize Population] $\rightarrow$ B[Compute Fitness using Equation (1)] $\rightarrow$ C[Select Parents] $\rightarrow$ D[Crossover] $\rightarrow$ E[Mutation] $\rightarrow$ F[Evaluate New Population] $\rightarrow$ G[Terminate Condition Met?]

[Yes] $\rightarrow$ H[Output Optimal Solution]

[No] $\rightarrow$ B

The genetic algorithm iteratively generates and evaluates a population of candidate solutions, using operations like selection, crossover, and mutation to explore the solution space and converge to the optimal solution.
\\3. By substituting the optimal solution $x^*$ into the objective function $f(x)$, we obtain the minimum total task delay $f(x^*)$, which constitutes the training data.
\subsection{Knowledge Distillation}
The simplified student model captures the feature extraction capabilities of both RNN and GNN through the encoder-decoder architecture to generate time series predictions.
Compared to the more complex teacher model, this student model achieves a trade-off between performance and complexity through model compression and knowledge distillation.
The simplified student model contains the following main components:
\\1. Encoder RNN module ($encoder\_rnn$): This module uses the RNNModel to encode the input rnninput.
\\2. Encoder GNN module ($encoder\_gnn$): This module uses the GNNModel to encode the input gnninput.
\\3.Decoder module (decoder): This linear layer module concatenates the outputs of the RNN and GNN encoders, and then outputs the final time series prediction.
The forward propagation process of the model is as follows:
\\1. First, create an empty graph $g$ with $num\_nodes$ nodes.
\\2. Input $gnninput$ into the $encoder\_gnn$ module to obtain $encoded\_gnn$.
\\3. Input $rnninput$ into the $encoder\_rnn$ module to obtain $rnn\_output$.
\\4. Copy and expand $rnn\_output$ to have the same shape as $encoded\_gnn$.
\\5. Concatenate $encoded\_gnn$ and the expanded $rnn\_output$ to obtain the final output output.
\\6. Input output into the decoder linear layer to obtain the final time series prediction $time\_sequence$.
\section{Experiments}\
The hardware platform used in this experiment was a server with an 13th Gen Intel(R) Core(TM) i9-13900H 2.60 GHz processor, 64GB RAM, and NVIDIA Geforce RTX 4090 GPU.
The software platform used was Python 3.7, with the following libraries and versions:
torch 2.1.0+cupy-cuda110 12.3.0,
torchvision 0.16.0,
torchreid 1.2.5,
Additionally, the following software versions were employed:
dgl 2.0.0+cu121,
numpy 1.26.2,
deap 1.4.1.
\subsection{Experiment design}\
To test the performance of this game model, we conducted a comprehensive evaluation of the model from the aspects of solution optimization, solution (game action) diversity, and the similarity between the solution and the optimal solution.
We designed a single-objective GA optimization algorithm to implement the meta strategy, generating a large number of task scenarios and their optimal solution labels as the training data for the deep learning model.
The labels are the action sets of the optimal resource allocation plans obtained by GA, and the data includes the task attribute vectors (the input of the GNN encoder) and the resource vectors of the three game participants (the input of the RNN encoder).
By training on a large number of sample data, the model learns the optimal game strategy.
The data is divided into training and test sets, each with 1000 samples.
The iteration number is 1000 in training phase, and the iteration number is 100 in testing phase.
Finally, in the network distillation system, we test the performance of the student model.
Through comparative experiments with GA, deep reinforcement learning, and DQN and MADDPG algorithms, we comprehensively evaluated the gaming effect of the model as a community-contribution-based decentralized consensus mechanism for blockchain.
\subsection{Performance evaluation}\
\begin{figure}[h]
\centering
\includegraphics[scale=0.43]{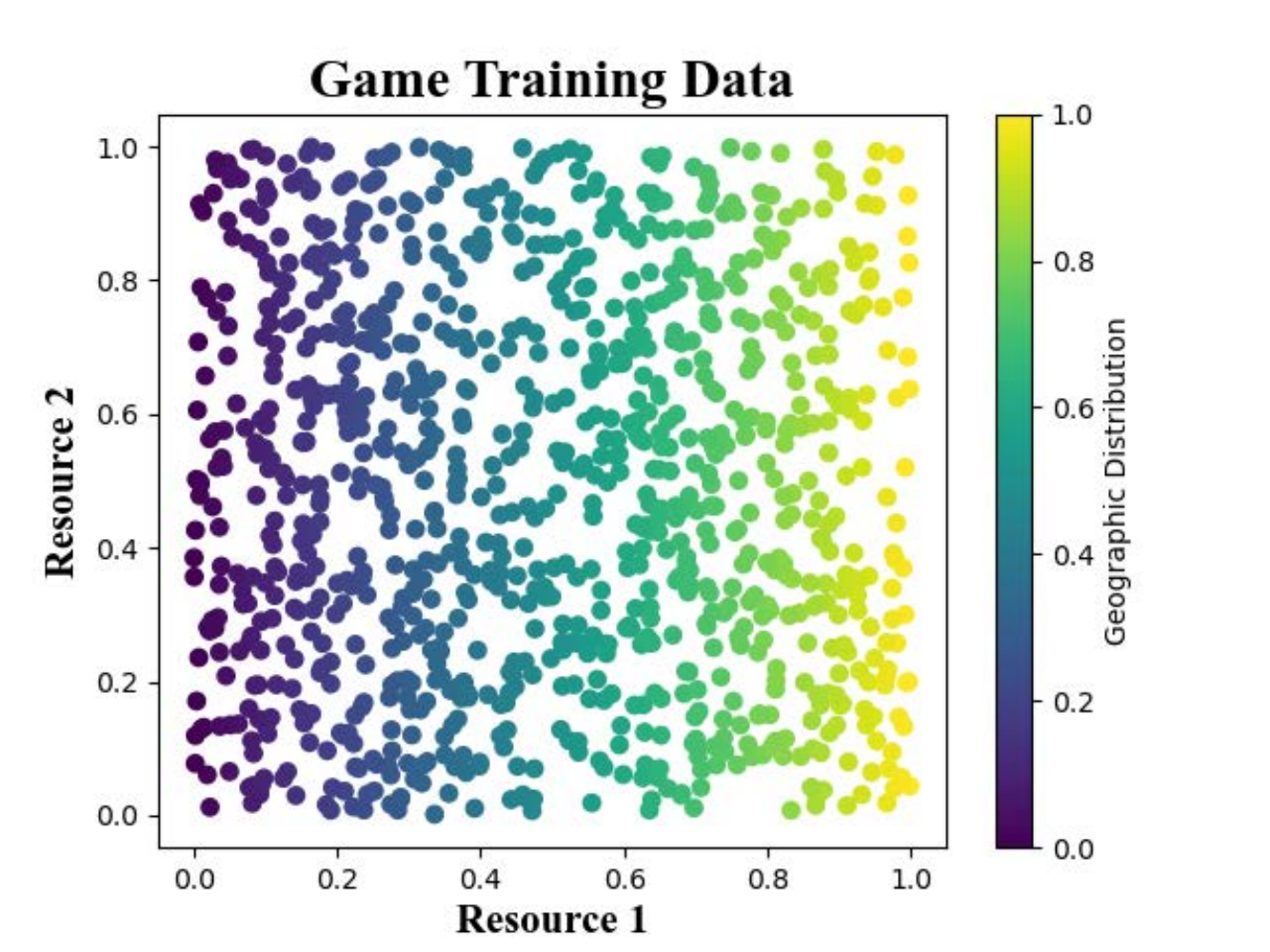}
\caption{Game training data}
\label{figure03}
\end{figure}
As shown in Figure \ref{figure03}, this is a visualization of the resource demand data of tasks in the generated game training data, where the x-axis and y-axis represent the demand quantities of different resource types, respectively.
The data in the figure is random and has been normalized.

\begin{figure}[h]
\centering
\includegraphics[scale=0.43]{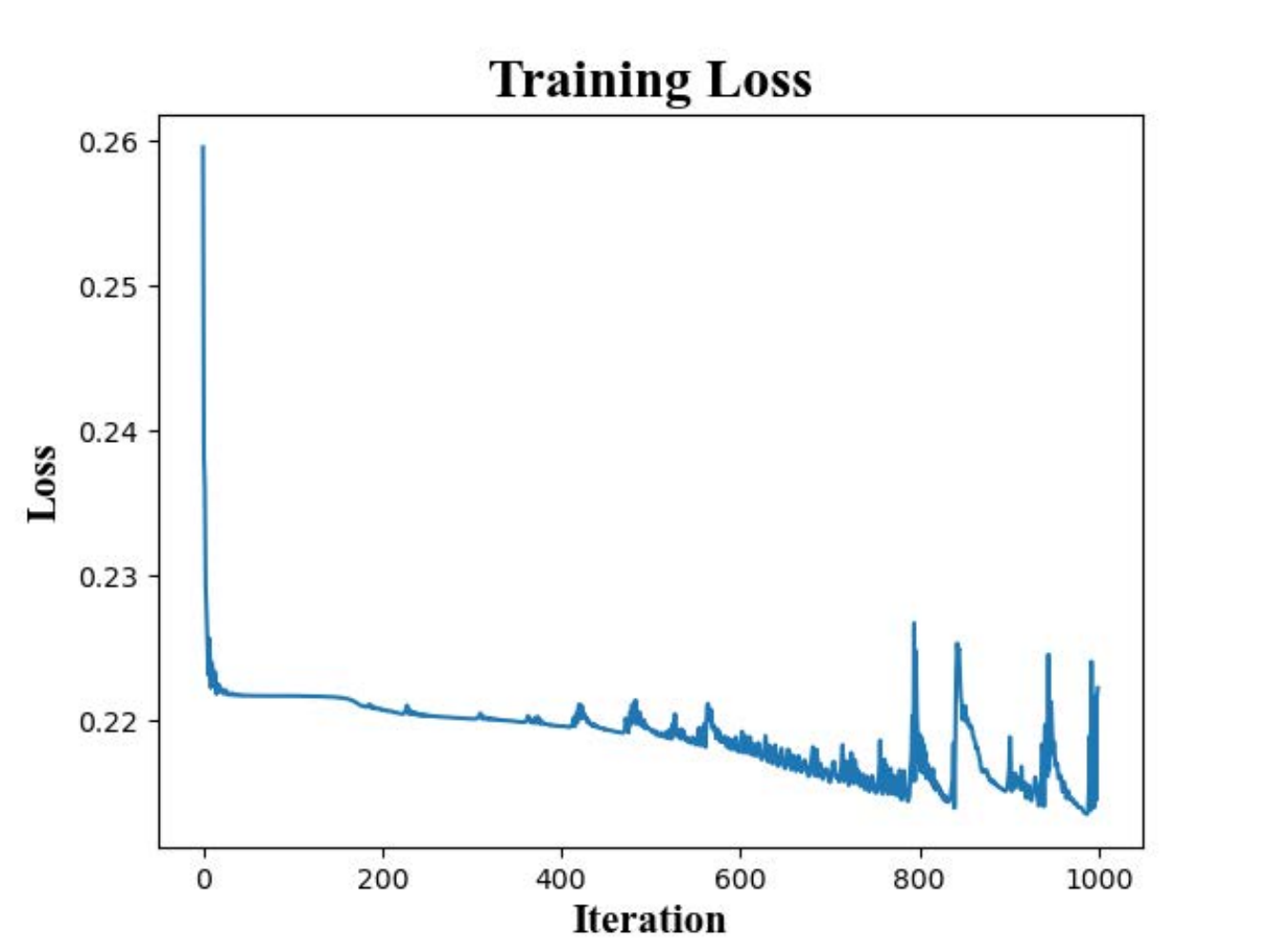}
\caption{Training Loss}
\label{figure04}
\end{figure}
Figure \ref{figure04} illustrates the evolution of the loss function during the training of the deep learning model.
The horizontal axis corresponds to the training epochs, while the vertical axis denotes the actual loss values.
As depicted, the initial loss value is relatively high, but it steadily declines as the training progresses.
This trend suggests that the model is effectively learning and improving its performance over time.
The consistent downward trajectory of the loss function indicates that the training is proceeding as expected, with the model converging towards an optimal solution.
Monitoring the loss function plot is crucial for evaluating the training process and identifying potential issues, such as overfitting or underfitting.
The decrease in the loss function, as observed in the figure, is a positive sign, signaling the successful training of the model.

\begin{figure}[h]
\centering
\includegraphics[scale=0.43]{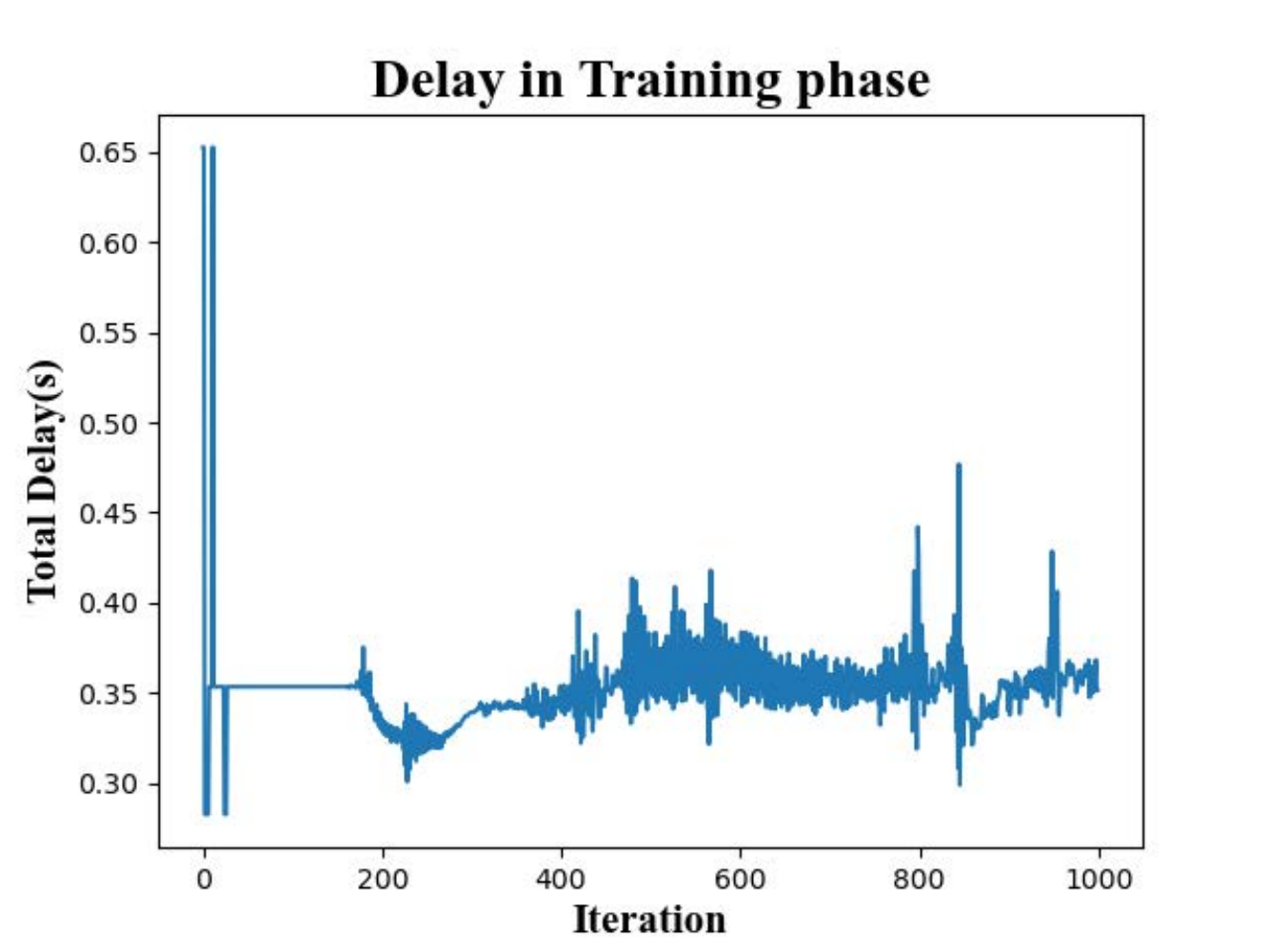}
\caption{Total delay in training phase}
\label{figure05}
\end{figure}
As shown in Figure\ref{figure05}, at the start of training, the model's parameters are randomly initialized, resulting in a relatively high target function value.
This indicates that the model is unable to fit the data well at the beginning of training. As training progresses, the model continuously adjusts its parameters, causing the target function value to gradually decrease.
This shows that the model is learning and improving its ability to fit the data.
After multiple training iterations, the target function value ultimately stabilizes and converges to a relatively small value.
This means that the model has learned the underlying patterns in the data and achieved a relatively optimal fitting effect.
The convergence of the target function minimization, as well as the balanced performance of the model on the training and validation sets, all suggest that the training process has been successful, and the model has learned the inherent laws of the data.
This lays a solid foundation for the subsequent application and deployment of the model.
But due to excessive training iterations, as observed in the figure, overfitting occurred near the 200th iteration.
Early stopping can be used to avoid overfitting.

\begin{figure}[h]
\centering
\includegraphics[scale=0.43]{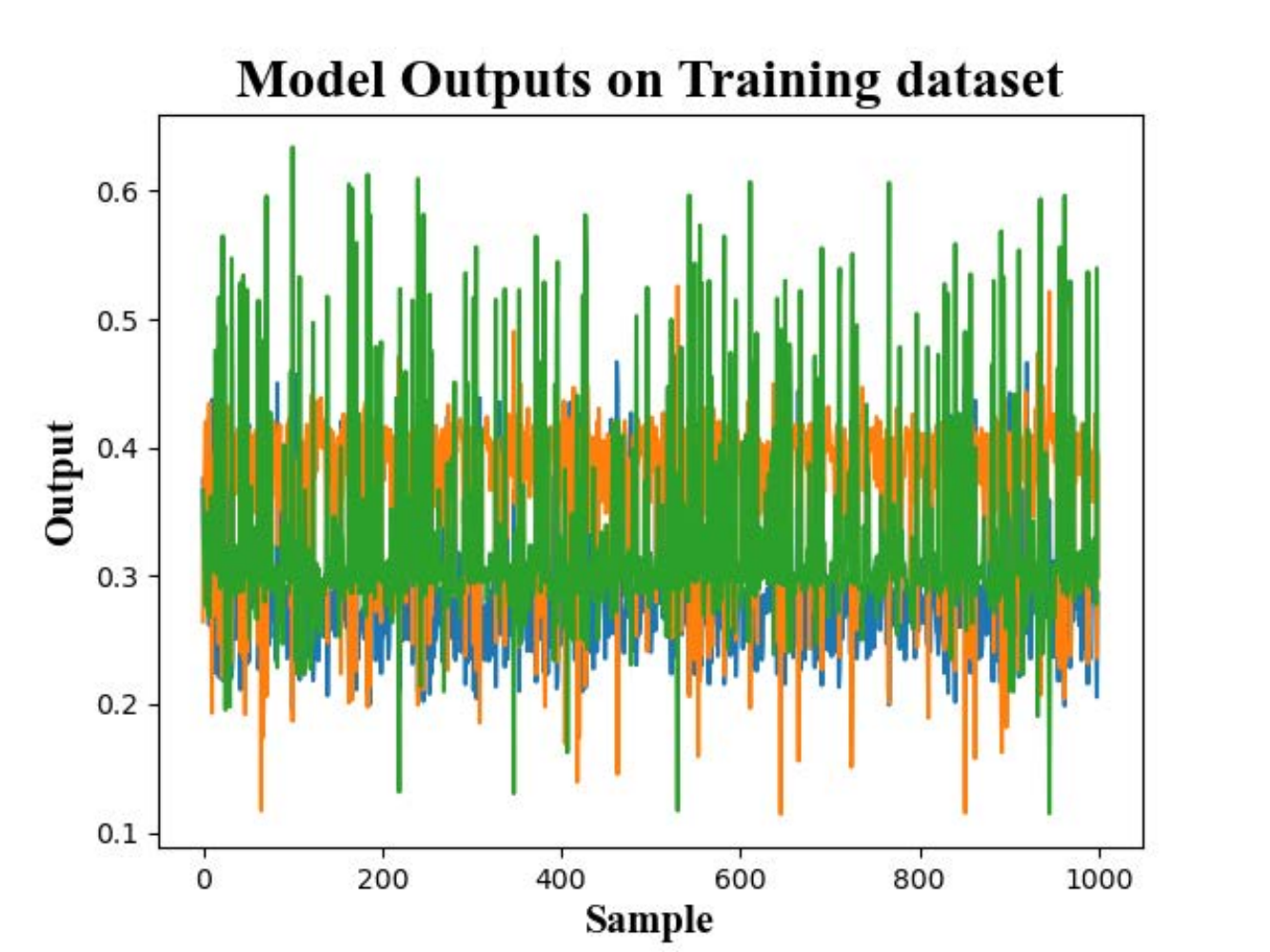}
\caption{Game model outputs in training phase}
\label{figure06}
\end{figure}
As shown in Figure \ref{figure06}, in the final iteration (100th iteration), are predictions of the players' game actions for 1000 task samples in training phase.
Each task is independently completed by one of the parties.
The three parties have reached a consensus on task allocation, with each party taking on different tasks, forming a clear division of labor and collaboration.
This ensures that the interests of all parties are fairly represented.
The outcome of the resource trading game is to form the consensus mechanism of the blockchain, contributing to the service quality of the community.

\begin{figure}[h]
\centering
\includegraphics[scale=0.43]{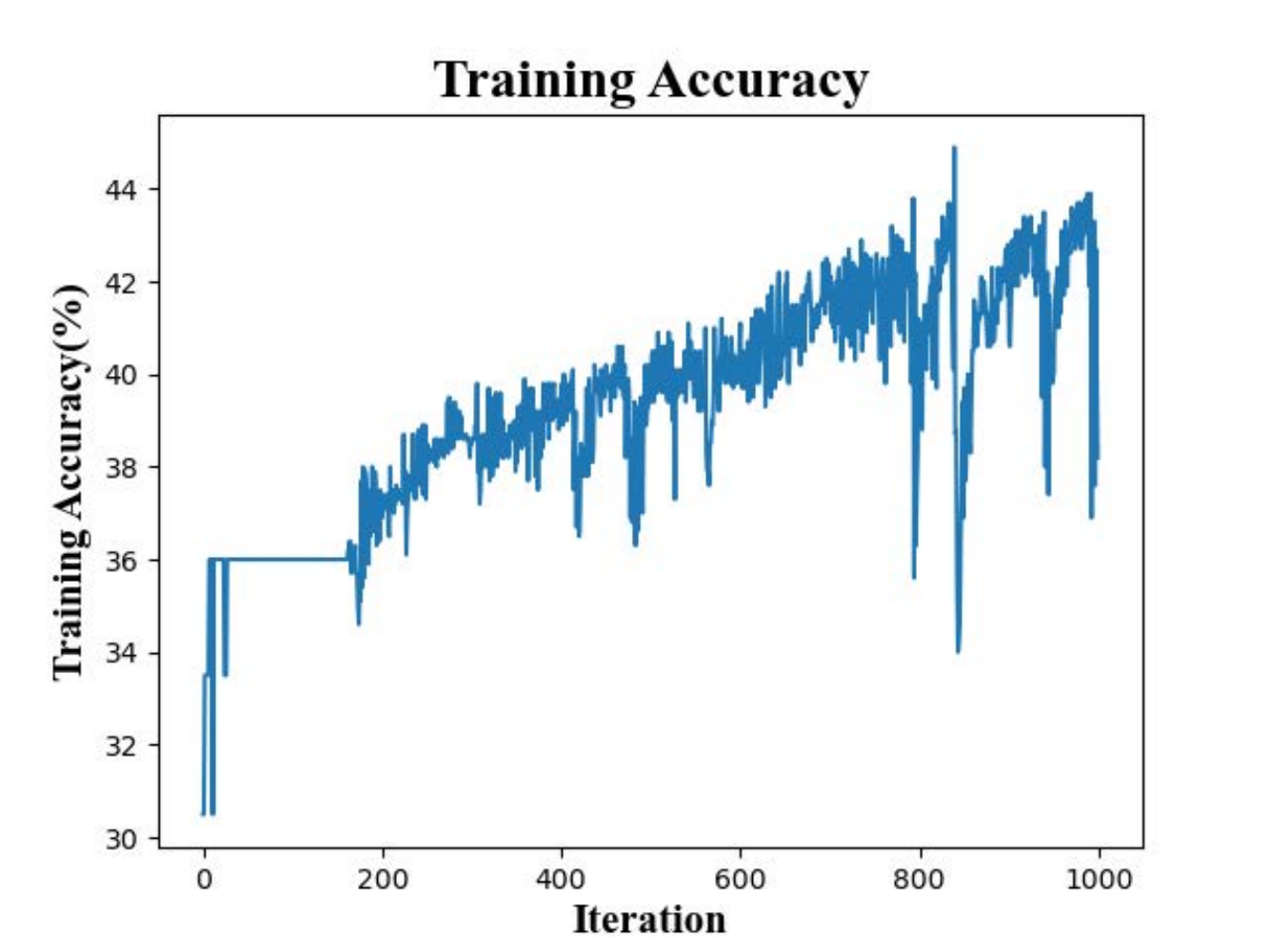}
\caption{Training Accuracy}
\label{figure07}
\end{figure}
We compared the similarity between the training results and the optimal solution generated by the meta-strategy.
The similarity accuracy between the resource allocation actions from the training results and the training data in each iteration cycle is shown in Figure \ref{figure07}.
As the number of iterations increases, the predicted values from the training become increasingly closer to the training data.
This indicates that the model is continuously adjusting its parameters and improving its ability to fit the data, gradually learning the underlying patterns in the sequential data.

\begin{figure}[h]
\centering
\includegraphics[scale=0.43]{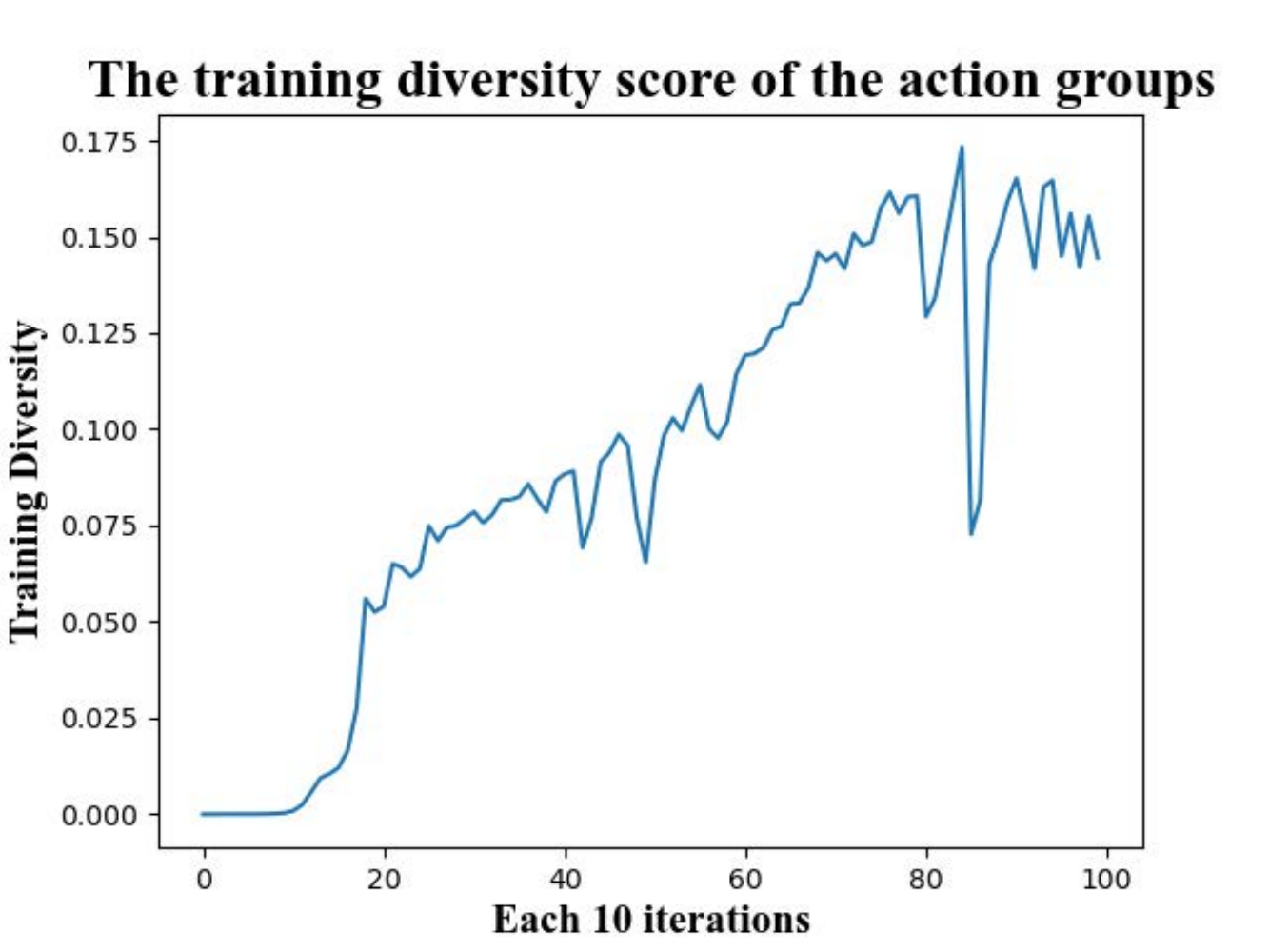}
\caption{Training diversity score}
\label{figure08}
\end{figure}
As shown in Figure \ref{figure07}, the diversity score (D) value increases, it indicates that the diversity among the action groups increases as the training samples increase, and the probability distribution differences between different action groups are relatively large.
In game theory, strategy diversity is an important metric for evaluating the effectiveness of a game model.
The increase in diversity score indicates that the game model has a richer repertoire of strategies to deal with different scenarios.
Since participants in the real world often adopt different strategies, the ability to adapt to various game situations can improve the model's robustness.
Strategy diversification can also prevent the model from getting stuck in local optima during optimization.
Furthermore, the obtained diversified results can enhance Byzantine fault tolerance and avoid being vulnerable to attacks due to fixed, known outcomes.
In summary, strategy diversity is crucial for the game model to perform well, adapt to complex environments, and avoid potential security risks.

\begin{figure}[h]
\centering
\includegraphics[scale=0.43]{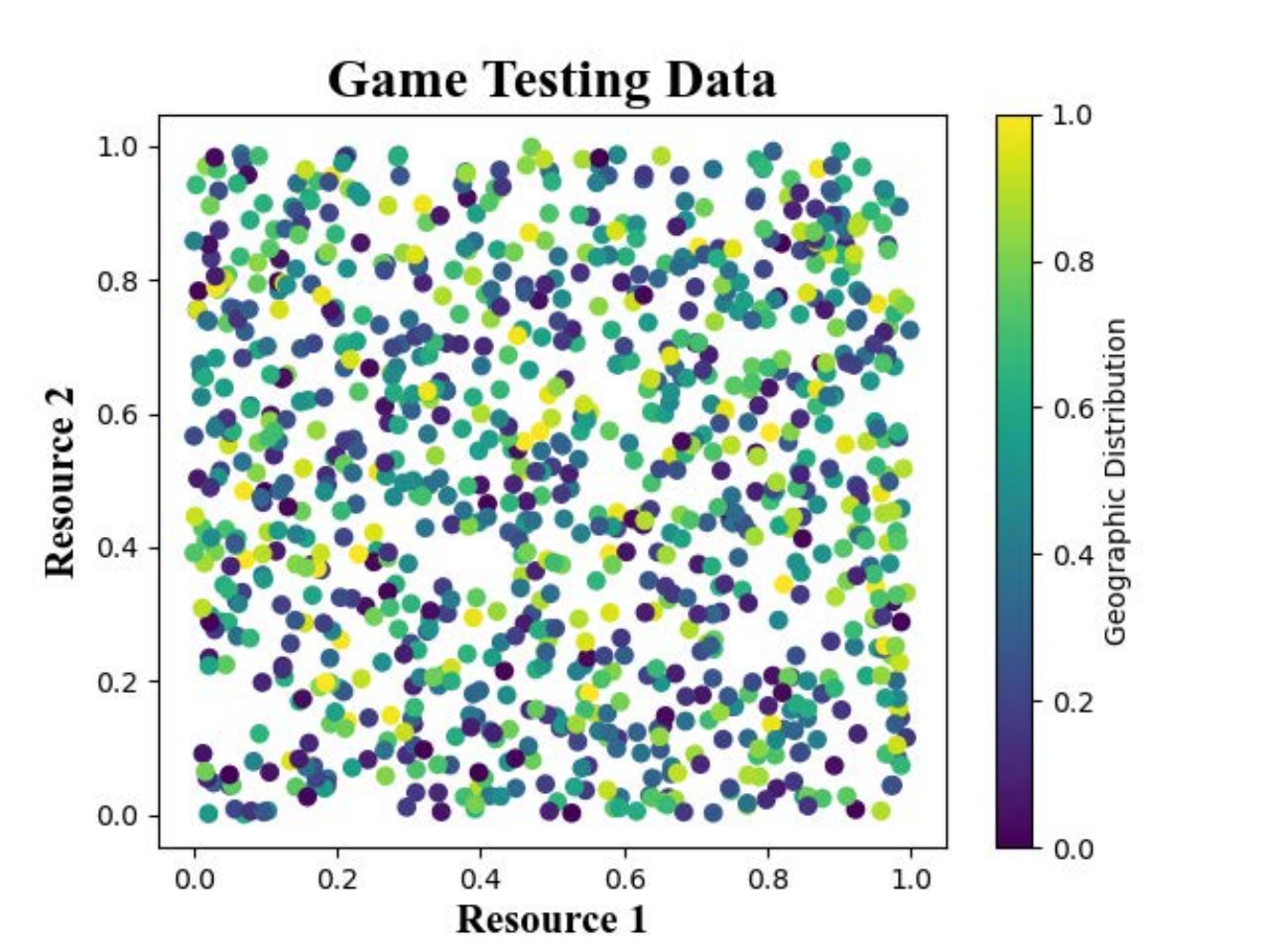}
\caption{Game testing data}
\label{figure09}
\end{figure}
As shown in Figure \ref{figure09}, this is a visualization of the resource demand data of tasks in the generated game testing data, where the x-axis and y-axis represent the demand quantities of different resource types, respectively.
The data in the figure is random and has been normalized.

\begin{figure}[h]
\centering
\includegraphics[scale=0.43]{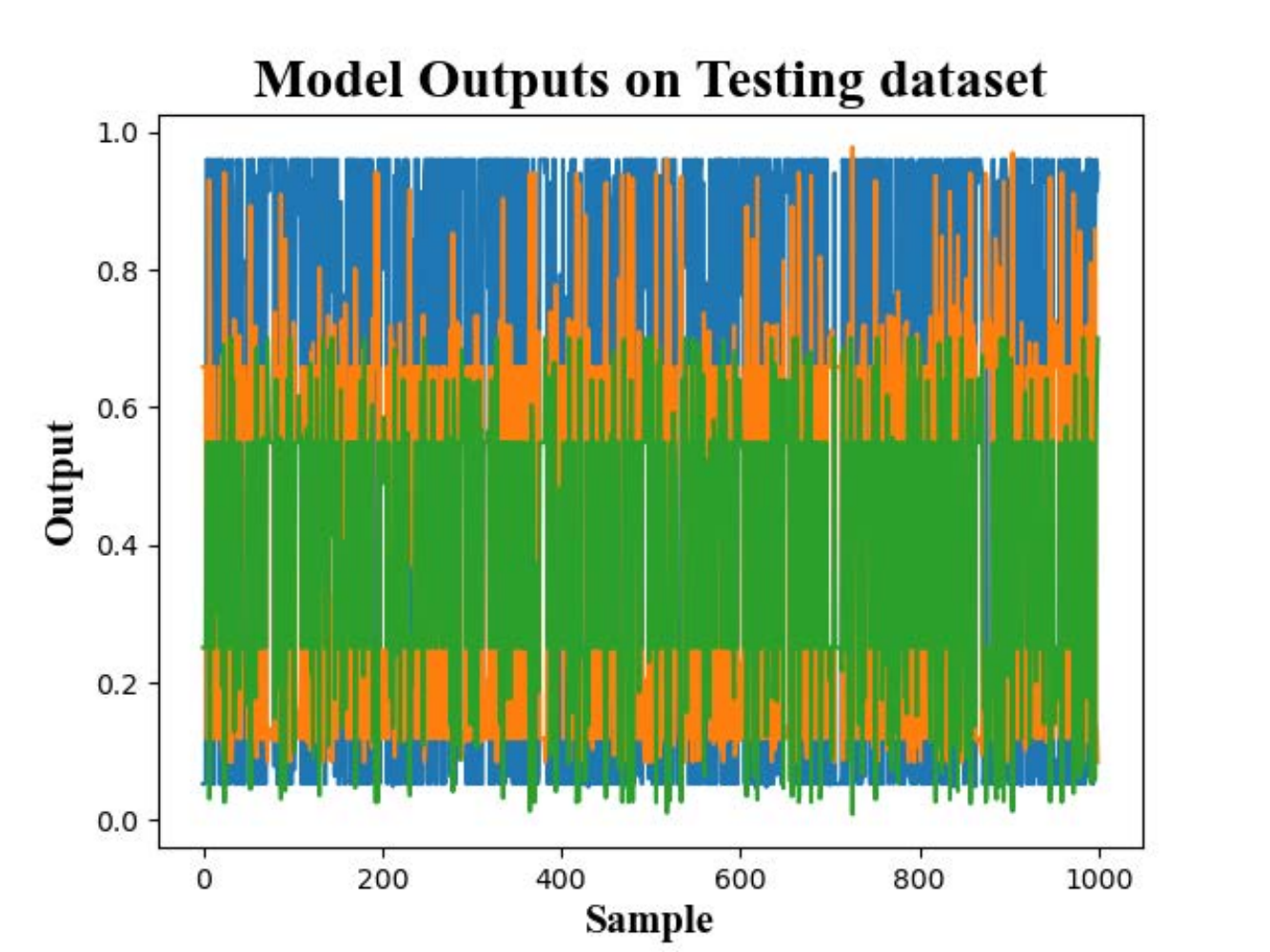}
\caption{Game model outputs in testing phase}
\label{figure10}
\end{figure}
As shown in Figure \ref{figure10}, in the final iteration (1000th iteration), are predictions of the players' game actions for 1000 task samples in testing phase.
The outcome of the resource trading game is to form the consensus mechanism of the blockchain, contributing to the service quality of the community.
The parties have reached a consensus on who provides the resources for the resource trading game.
For each task, the resource provider contributes the resource trading to the blockchain.
This collaborative game not only ensures the fair representation of the interests of all parties, but also enables the edge community to provide low-latency services.

\begin{figure}[h]
\centering
\includegraphics[scale=0.43]{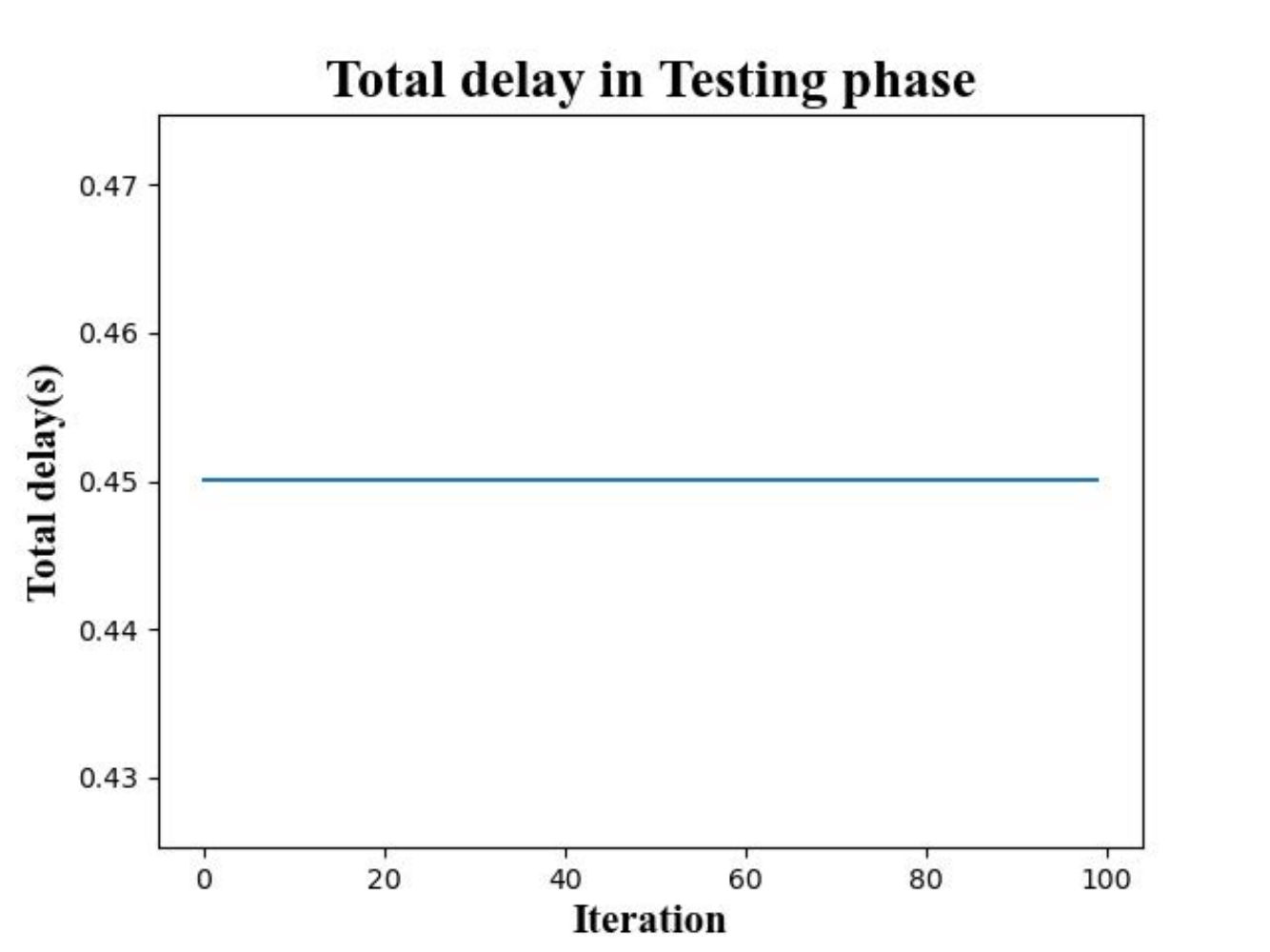}
\caption{Total delay in testing phase}
\label{figure11}
\end{figure}

\begin{figure}[h]
\centering
\includegraphics[scale=0.43]{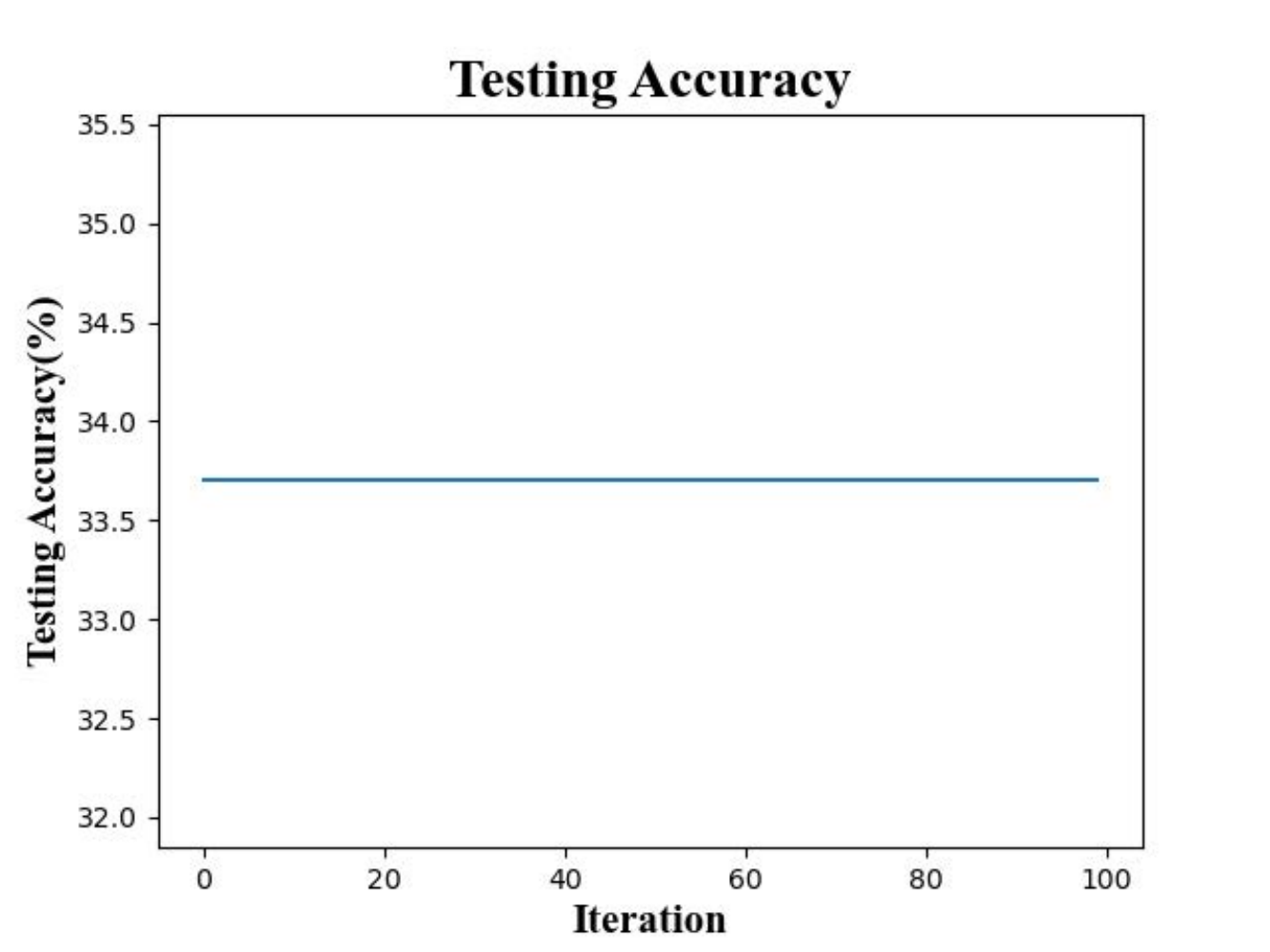}
\caption{Testing accuracy}
\label{figure12}
\end{figure}

\begin{figure}[h]
\centering
\includegraphics[scale=0.43]{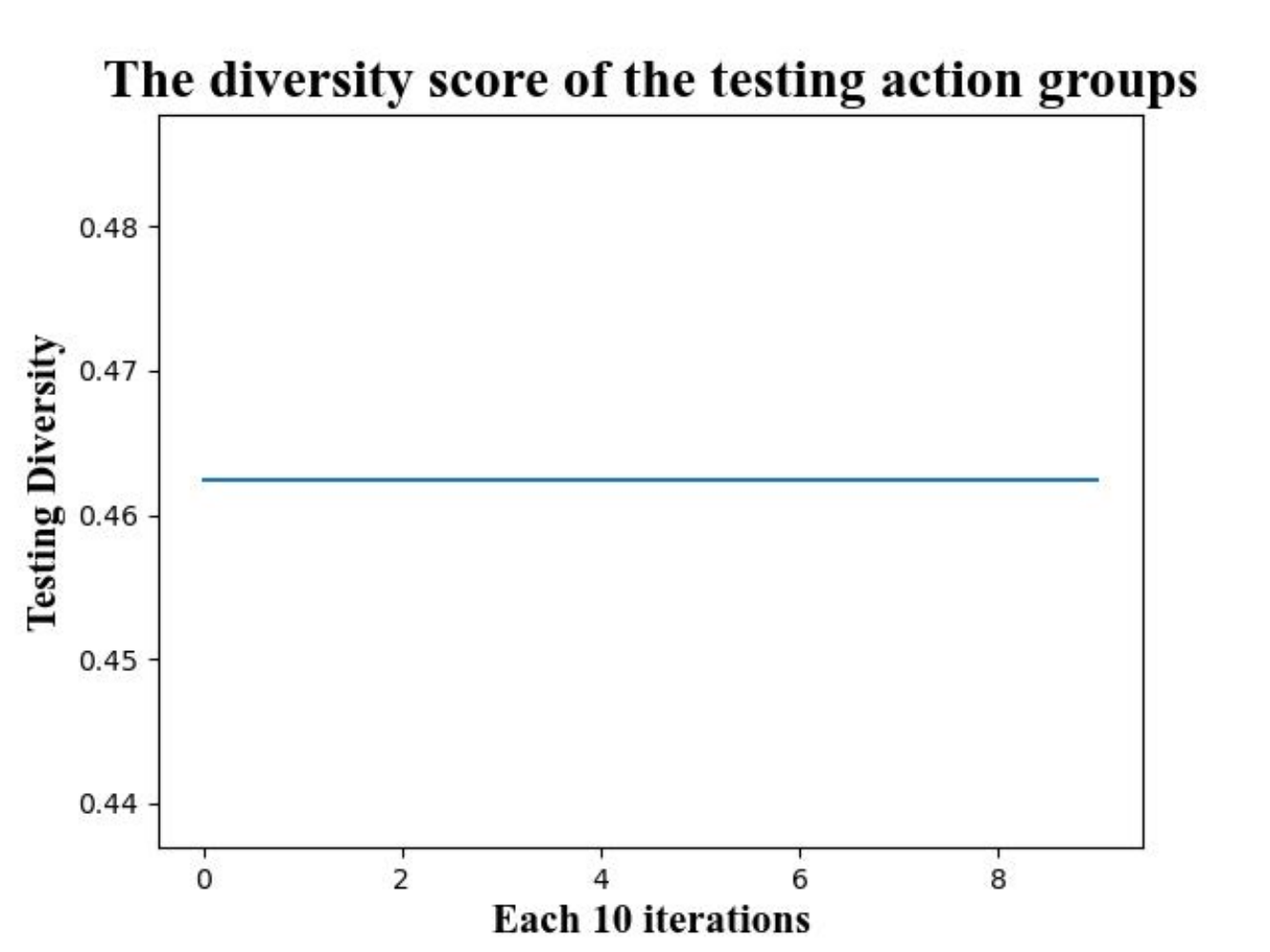}
\caption{Testing diversity score}
\label{figure13}
\end{figure}
As shown in Figure \ref{figure11} - Figure \ref{figure13}, during the entire iteration cycle of the testing, the model parameters remain unchanged, so the test data obtained is also constant.
The results on the validation set demonstrate the model's performance on new data, and the validation set performance is on par with the training set.
Compared to the graph from the training stage, the above three graphs taken together indicate that the model has very good generalization ability.
Deep learning models are typically non-convex optimization problems, and there exist multiple local optimal solutions.
The training process of deep learning adopts stochastic gradient descent, and the randomness of initial parameters as well as random sampling of the dataset both affect the optimization results, leading to different local optimal solutions.
Therefore, the multiple different optimal solutions and the diversity of scores for each solution can enhance the Byzantine fault tolerance of the entire game consensus system, improving the robustness of the PoC system.

\begin{figure}
\centering
\includegraphics[scale=0.43]{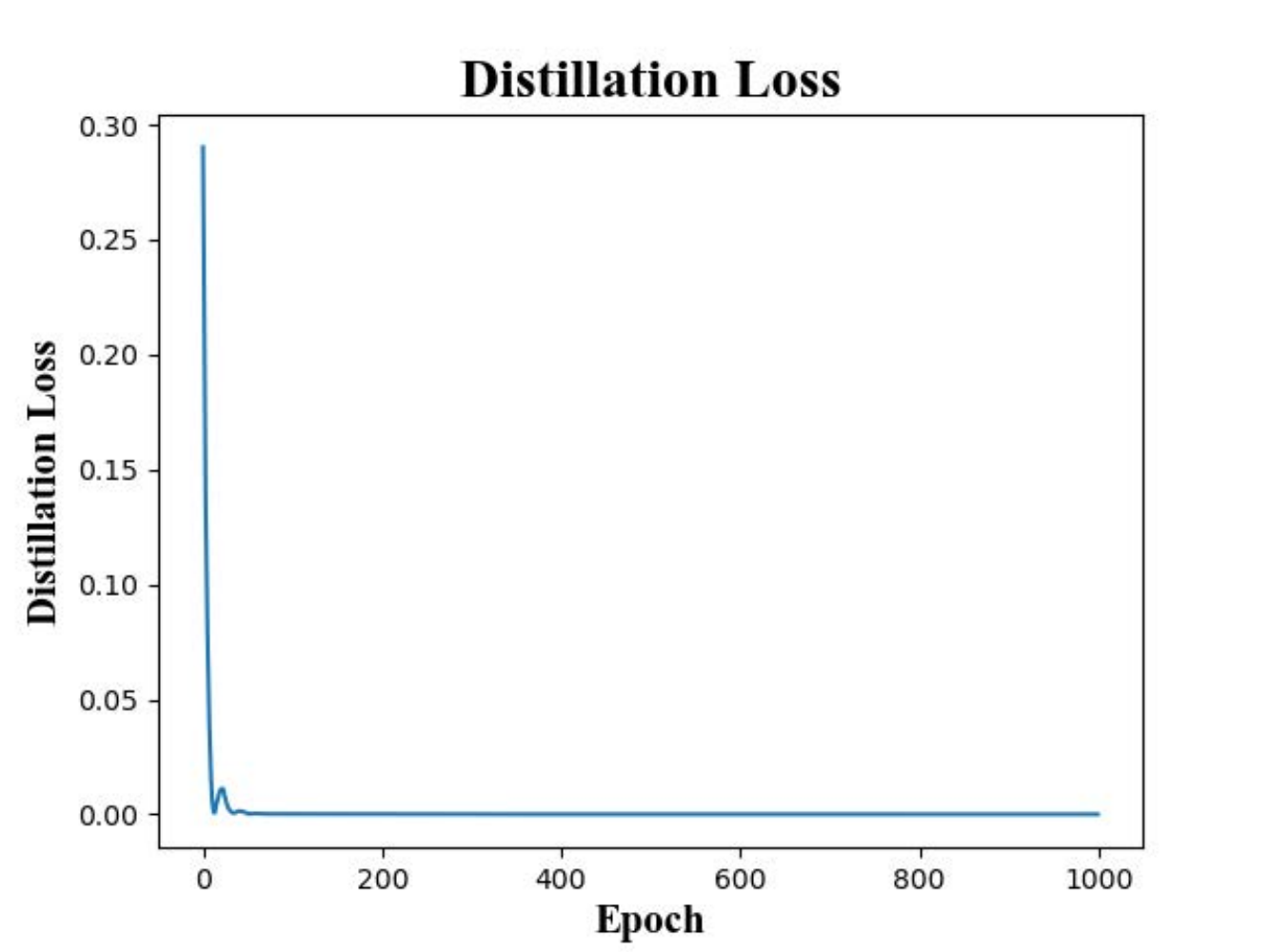}
\caption{Distillation Loss}
\label{figure14}
\end{figure}

\begin{figure}
\centering
\includegraphics[scale=0.43]{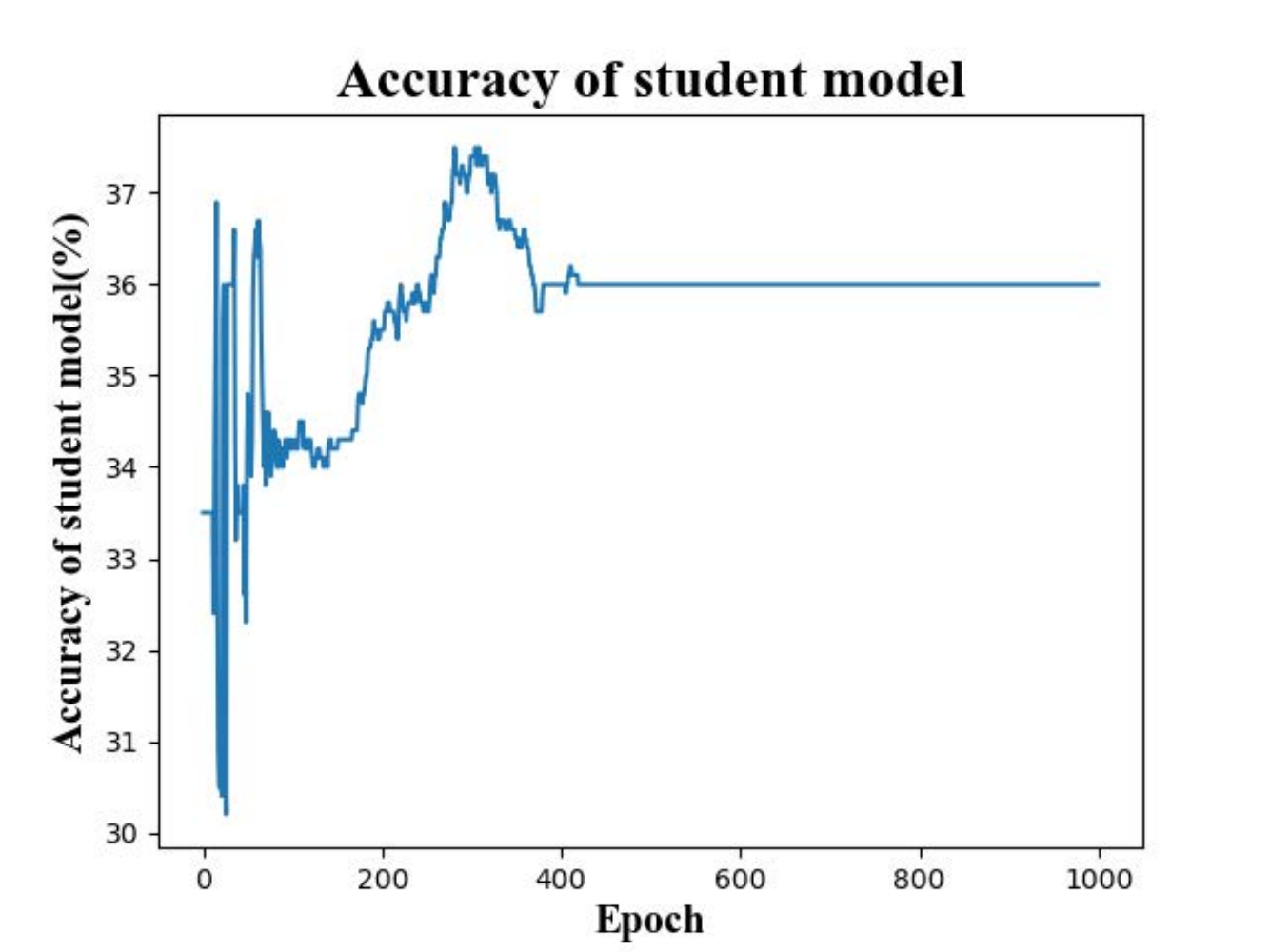}
\caption{Accuracy of student model}
\label{figure15}
\end{figure}

\begin{figure}
\centering
\includegraphics[scale=0.43]{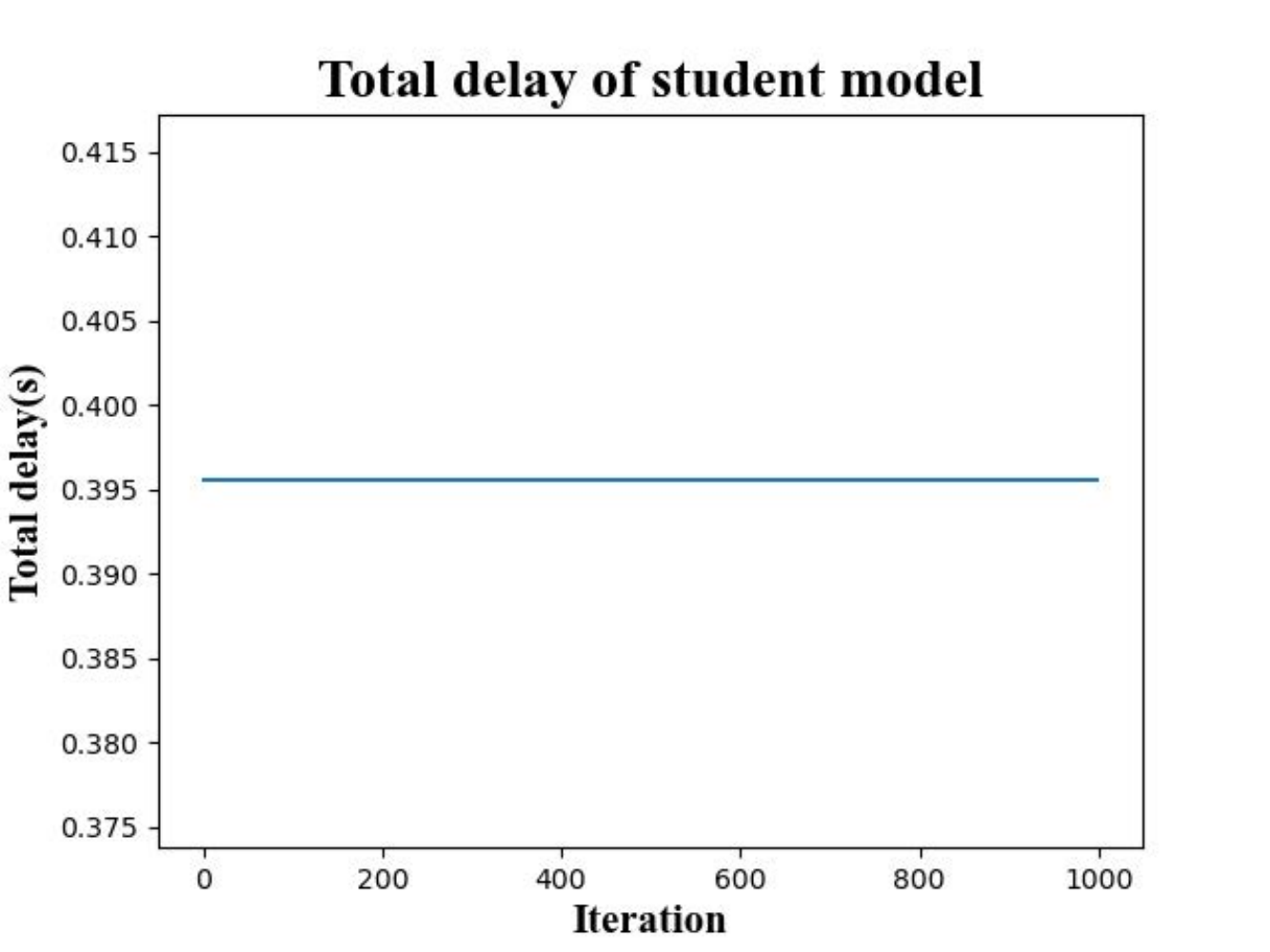}
\caption{Total delay of student model}
\label{figure16}
\end{figure}
As illustrated by Figure \ref{figure14} - Figure \ref{figure15}, in the student model training of the knowledge distillation system, as the loss decreases, the model parameters gradually converge, and can effectively learn the knowledge of the teacher model.
Through reasonable optimization algorithms and hyperparameter adjustments, the performance of the student model is improved.
The student model's parameters are continuously optimized and eventually stabilize, realizing the knowledge transfer from the teacher model to the student model.
As illustrated by Figure \ref{figure16}, the teacher model outperforms the student model by $19.37\%$.
This is because the teacher model is trained on the complete original data, while the student model is trained by imitating the knowledge of the teacher model.
By increasing the diversity and scale of the training data or introducing more types of input features, the student model can have more optimization space.

\begin{figure}
\centering
\includegraphics[scale=0.43]{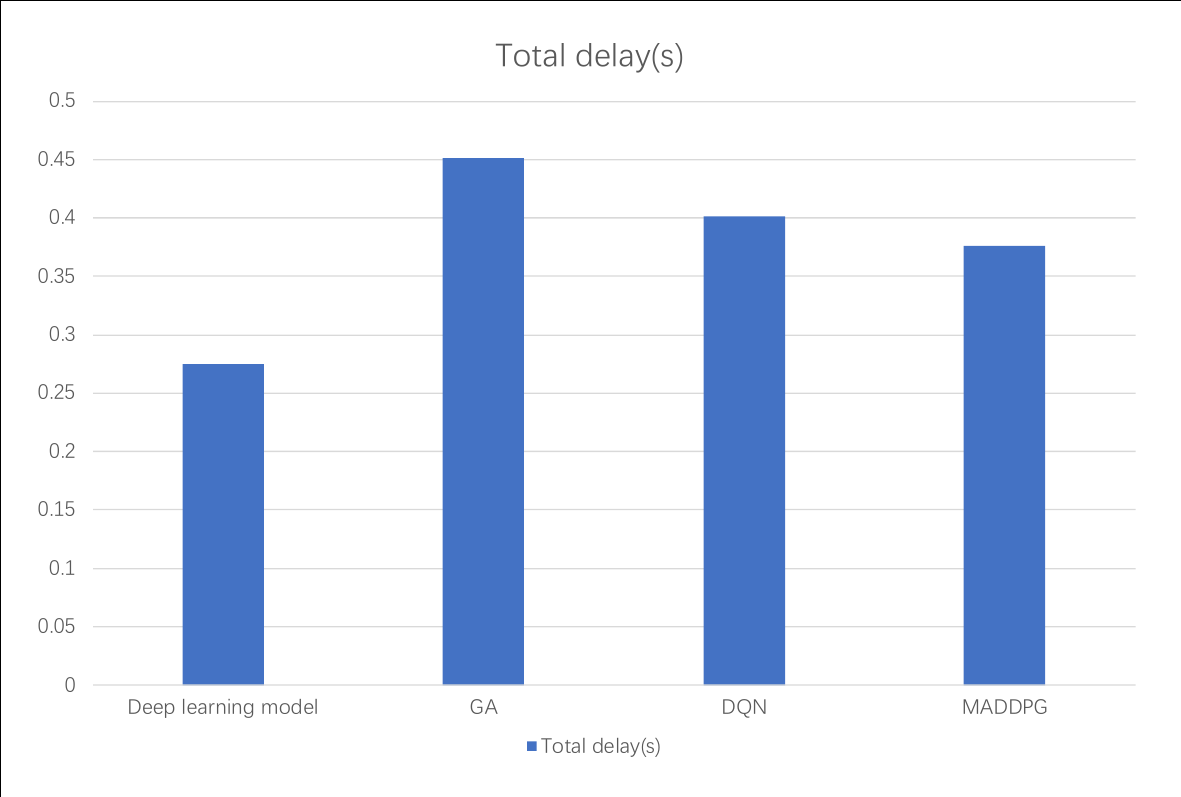}
\caption{Algorithms comparison}
\label{figure17}
\end{figure}
As illustrated by Figure \ref{figure17}, the Deep model, although using training data generated by GA, performs better than GA in optimization exploration.
This indicates that the Deep learning model has already mastered the underlying rules of the resource allocation plan sequence, and the model has strong representation capabilities.
When the service delay objective function is relatively smooth and differentiable, DQN (Deep Q-Network) and MADDPG can typically converge faster and achieve better performance than GA.
In complex multi-agent environments, MADDPG is likely to produce better optimization solutions compared to DQN.
MADDPG is more suitable for multi-agent environments, as it can learn effective policies in the presence of other learning agents.
The proposed deep learning model has achieved a performance improvement of over $5.84\%$ compared to the state-of-the-art algorithms.
\section{Conclusion}\
In edge computing, horizontal collaboration is becoming increasingly important.
The optimization and collaborative game of resources from different sites, and even different service providers, can effectively solve the problem of limited and uneven distribution of edge resources, which leads to service constraints.
In the large-scale deployment of edge nodes, maintaining blockchain records in the peer-to-peer network of edge nodes is very necessary due to considerations of decentralization and credibility.
The blockchain consensus mechanism used in previous work cannot guarantee the low latency in the edge network.
The proposed contribution-based proof of consensus mechanism can be beneficial to the service quality of the edge community.
Due to the natural connection between deep learning networks and game theory, the proposed dual encoder-single decoder deep model learning model can effectively learn and simulate the game process of the blockchain consensus mechanism based on the meta-strategy of dual objectives.
Extensive experiments have proven that using the optimized solution generated by the meta-strategy as training data can simulate game-theoretic behavior of the participating nodes and avoid a single stakeholder controlling the entire system.
The entire consensus mechanism is built on the foundation of the game participants' contributions to community services, which is beneficial for low-latency services in the edge network.
In future work, we will further include the solutions of resource allocation problems of different scales and complexities in the training samples to enhance the generalization capability of the model.
Multi-objective learning is also a research direction for our future expansions.
\appendices
%

%
%
%

\ifCLASSOPTIONcaptionsoff
  \newpage
\fi



%

%
\bibliographystyle{IEEEtran}
\bibliography{RefJWang}

\begin{thebibliography}{10}
\providecommand{\url}[1]{#1}
\csname url@samestyle\endcsname
\providecommand{\newblock}{\relax}
\providecommand{\bibinfo}[2]{#2}
\providecommand{\BIBentrySTDinterwordspacing}{\spaceskip=0pt\relax}
\providecommand{\BIBentryALTinterwordstretchfactor}{4}
\providecommand{\BIBentryALTinterwordspacing}{\spaceskip=\fontdimen2\font plus
\BIBentryALTinterwordstretchfactor\fontdimen3\font minus
  \fontdimen4\font\relax}
\providecommand{\BIBforeignlanguage}[2]{{%
\expandafter\ifx\csname l@#1\endcsname\relax
\typeout{** WARNING: IEEEtran.bst: No hyphenation pattern has been}%
\typeout{** loaded for the language `#1'. Using the pattern for}%
\typeout{** the default language instead.}%
\else
\language=\csname l@#1\endcsname
\fi
#2}}
\providecommand{\BIBdecl}{\relax}
\BIBdecl

\bibitem{f4}
M.~Zhang, J.~Cao, Y.~Sahni, Q.~Chen, S.~Jiang, and L.~Yang, ``Blockchain-based
  collaborative edge intelligence for trustworthy and real-time video
  surveillance,'' \emph{IEEE Transactions on Industrial Informatics}, vol.~19,
  no.~2, pp. 1623--1633, 2022.

\bibitem{f1}
T.~Hazra and K.~Anjaria, ``Applications of game theory in deep learning: a
  survey,'' \emph{Multimedia Tools and Applications}, vol.~81, no.~6, pp.
  8963--8994, 2022.

\bibitem{f7}
X.~Li, P.~Jiang, T.~Chen, X.~Luo, and Q.~Wen, ``A survey on the security of
  blockchain systems,'' \emph{Future generation computer systems}, vol. 107,
  pp. 841--853, 2020.

\bibitem{f8}
M.~Ghosh, M.~Richardson, B.~Ford, and R.~Jansen, ``A torpath to torcoin,
  proof-of-bandwidth altcoins for compensating relays (2014),'' \emph{URL
  https://www. smithandcrown.
  com/open-research/a-torpath-to-torcoin-proof-of-bandw
  idth-altcoins-for-compensating-relays}, 2021.

\bibitem{f9}
S.~Joshi, ``Feasibility of proof of authority as a consensus protocol model,''
  \emph{arXiv preprint arXiv:2109.02480}, 2021.

\bibitem{f10}
L.~Lamport, R.~Shostak, and M.~Pease, ``The byzantine generals problem,''
  \emph{Concurrency: the works of leslie lamport}, pp. 203--226, 2019.

\bibitem{f11}
M.~H. Manshaei, M.~Jadliwala, A.~Maiti, and M.~Fooladgar, ``A game-theoretic
  analysis of shard-based permissionless blockchains,'' \emph{IEEE Access},
  vol.~6, pp. 78\,100--78\,112, 2018.

\bibitem{f6}
H.~Tembine, ``Deep learning meets game theory: Bregman-based algorithms for
  interactive deep generative adversarial networks,'' \emph{IEEE transactions
  on cybernetics}, vol.~50, no.~3, pp. 1132--1145, 2019.

\bibitem{f12}
D.~Silver, A.~Huang, C.~J. Maddison, A.~Guez, L.~Sifre, G.~Van Den~Driessche,
  J.~Schrittwieser, I.~Antonoglou, V.~Panneershelvam, M.~Lanctot \emph{et~al.},
  ``Mastering the game of go with deep neural networks and tree search,''
  \emph{nature}, vol. 529, no. 7587, pp. 484--489, 2016.

\bibitem{f13}
D.~Silver, J.~Schrittwieser, K.~Simonyan, I.~Antonoglou, A.~Huang, A.~Guez,
  T.~Hubert, L.~Baker, M.~Lai, A.~Bolton \emph{et~al.}, ``Mastering the game of
  go without human knowledge,'' \emph{nature}, vol. 550, no. 7676, pp.
  354--359, 2017.

\bibitem{f14}
O.~Vinyals, I.~Babuschkin, W.~M. Czarnecki, M.~Mathieu, A.~Dudzik, J.~Chung,
  D.~H. Choi, R.~Powell, T.~Ewalds, P.~Georgiev \emph{et~al.}, ``Grandmaster
  level in starcraft ii using multi-agent reinforcement learning,''
  \emph{Nature}, vol. 575, no. 7782, pp. 350--354, 2019.

\bibitem{f15}
G.~Dulac-Arnold, D.~Mankowitz, and T.~Hester, ``Challenges of real-world
  reinforcement learning,'' \emph{arXiv preprint arXiv:1904.12901}, 2019.

\bibitem{f16}
L.~Canese, G.~C. Cardarilli, L.~Di~Nunzio, R.~Fazzolari, D.~Giardino, M.~Re,
  and S.~Span{\`o}, ``Multi-agent reinforcement learning: A review of
  challenges and applications,'' \emph{Applied Sciences}, vol.~11, no.~11, p.
  4948, 2021.

\bibitem{f2}
H.~Lin, L.~Yang, H.~Guo, and J.~Cao, ``Decentralized task offloading in edge
  computing: An offline-to-online reinforcement learning approach,'' \emph{IEEE
  Transactions on Computers}, 2024.

\bibitem{f32}
J.~Gou, B.~Yu, S.~J. Maybank, and D.~Tao, ``Knowledge distillation: A survey,''
  \emph{International Journal of Computer Vision}, vol. 129, no.~6, pp.
  1789--1819, 2021.

\bibitem{f33}
Z.~Gao, K.~Xu, B.~Ding, and H.~Wang, ``Knowru: Knowledge reuse via knowledge
  distillation in multi-agent reinforcement learning,'' \emph{Entropy},
  vol.~23, no.~8, p. 1043, 2021.

\bibitem{f34}
G.~Hinton, O.~Vinyals, and J.~Dean, ``Distilling the knowledge in a neural
  network,'' \emph{arXiv preprint arXiv:1503.02531}, 2015.

\bibitem{f28}
Y.~Liu and M.~Lapata, ``Hierarchical transformers for multi-document
  summarization,'' \emph{arXiv preprint arXiv:1905.13164}, 2019.

\bibitem{f29}
Y.~Li, Z.~Wang, L.~Yin, Z.~Zhu, G.~Qi, and Y.~Liu, ``X-net: a dual
  encoding--decoding method in medical image segmentation,'' \emph{The Visual
  Computer}, pp. 1--11, 2023.

\bibitem{f30}
R.~Novak, M.~Auli, and D.~Grangier, ``Iterative refinement for machine
  translation,'' \emph{arXiv preprint arXiv:1610.06602}, 2016.

\bibitem{f31}
K.~Yao, L.~Zhang, D.~Du, T.~Luo, L.~Tao, and Y.~Wu, ``Dual encoding for
  abstractive text summarization,'' \emph{IEEE transactions on cybernetics},
  vol.~50, no.~3, pp. 985--996, 2018.

\bibitem{f17}
J.~Bergstra and Y.~Bengio, ``Random search for hyper-parameter optimization.''
  \emph{Journal of machine learning research}, vol.~13, no.~2, 2012.

\bibitem{f18}
J.~Snoek, H.~Larochelle, and R.~P. Adams, ``Practical bayesian optimization of
  machine learning algorithms,'' \emph{Advances in neural information
  processing systems}, vol.~25, 2012.

\bibitem{f19}
L.~Li, K.~Jamieson, G.~DeSalvo, A.~Rostamizadeh, and A.~Talwalkar, ``Hyperband:
  A novel bandit-based approach to hyperparameter optimization,'' \emph{Journal
  of Machine Learning Research}, vol.~18, no. 185, pp. 1--52, 2018.

\bibitem{f20}
F.~Chen, N.~Wang, J.~Tang, P.~Yan, and J.~Yu, ``Unsupervised person
  re-identification via multi-domain joint learning,'' \emph{Pattern
  Recognition}, vol. 138, p. 109369, 2023.

\bibitem{f21}
J.~Wang and C.~Song, ``Quantum learning and essential cognition under the
  traction of meta-characteristics in an open world,'' \emph{arXiv preprint
  arXiv:2311.13335}, 2023.

\bibitem{f22}
T.~Elsken, J.~H. Metzen, and F.~Hutter, ``Neural architecture search: A
  survey,'' \emph{Journal of Machine Learning Research}, vol.~20, no.~55, pp.
  1--21, 2019.

\bibitem{f23}
H.~Pham, M.~Guan, B.~Zoph, Q.~Le, and J.~Dean, ``Efficient neural architecture
  search via parameters sharing,'' \emph{International conference on machine
  learning}, pp. 4095--4104, 2018.

\bibitem{f24}
H.~Liu, K.~Simonyan, and Y.~Yang, ``Darts: Differentiable architecture
  search,'' \emph{arXiv preprint arXiv:1806.09055}, 2018.

\bibitem{f25}
J.~Friedman, T.~Hastie, and R.~Tibshirani, ``Additive logistic regression: a
  statistical view of boosting (with discussion and a rejoinder by the
  authors),'' \emph{The annals of statistics}, vol.~28, no.~2, pp. 337--407,
  2000.

\bibitem{f26}
L.~Breiman, ``Bagging predictors,'' \emph{Machine learning}, vol.~24, pp.
  123--140, 1996.

\bibitem{f27}
D.~H. Wolpert, ``Stacked generalization,'' \emph{Neural networks}, vol.~5,
  no.~2, pp. 241--259, 1992.

\end{thebibliography}
\end{document}